%% file: main.tex
\documentclass[conference,compsoc]{IEEEtran}
\AtBeginDocument{%
  }

\ifCLASSOPTIONcompsoc
  \usepackage[nocompress]{cite}
\else
  \usepackage{cite}
\fi

\usepackage{subfig}
\usepackage{listings}
\usepackage{multirow}
\usepackage{fancybox}
\usepackage{hyperref}
\usepackage{xurl}
\usepackage{amsmath,amssymb,amsfonts}
\usepackage{algorithmic}
\usepackage{graphicx}
\usepackage{tcolorbox}
\usepackage{textcomp}
\usepackage{booktabs}
\usepackage{xcolor}
\usepackage{float}
\usepackage{placeins}
\usepackage{needspace}
\usepackage{dblfloatfix}
\usepackage{mdframed}
\usepackage{soul} 
\usepackage{booktabs}
\usepackage{array}
\usepackage{tabularx}
\usepackage{makecell}
\usepackage{amssymb}

\ifCLASSINFOpdf
\else
\fi                                        
\begin{document}

\title{``I Don't Think It's a Scam'': A First Look at Family Intervention Barriers in Elder Abuse Discussions on Reddit and Baidu Tieba}

\author{
\IEEEauthorblockN{Yipeng Wang\IEEEauthorrefmark{2},
Huy Gia Han Vu$^\diamond$\IEEEauthorrefmark{1},
Jiaming Liu\IEEEauthorrefmark{2},
Mihika Mehta\IEEEauthorrefmark{2}, \\
Poojitha Thota\IEEEauthorrefmark{3}, 
Shirin Nilizadeh\IEEEauthorrefmark{4},
Mohit Singhal\IEEEauthorrefmark{2}}
\IEEEauthorblockA{\IEEEauthorrefmark{2} Northeastern University \\
\IEEEauthorblockA{\IEEEauthorrefmark{1} Rutgers University-Newark 
\IEEEauthorblockA{\IEEEauthorrefmark{3} Rockwell Automation Inc. \\
\IEEEauthorblockA{\IEEEauthorrefmark{4} The University of Texas at Arlington \\
(wang.yipen,liu.jiamin,mehta.mihik,m.singhal)@northeastern.edu \\ han.vu@rutgers.edu, Poojitha.Thota@rockwellautomation.com, shirin.nilizadeh@uta.edu
}}}}}

\maketitle
\def\thefootnote{$\diamond$}\footnotetext{Work done at Northeastern University}
\begin{abstract} 
When older adults trust scammers despite warnings, families face obstacles to intervention. Complainants aged 60 and older reported \$4.9 billion in losses in 2024 to the U.S. Internet Crime Complaint Center. To investigate family intervention barriers to elder scams, we collected 2M Reddit posts from May 2021 to June 2026 and 2B Baidu Tieba posts from 2007 to 2026. The final corpora contain 13,082 Reddit posts and 3,933 Baidu Tieba posts, labeled using platform-specific codebooks, manual reference annotations, and LLM coding. Through a mixed-methods analysis, we find that scams often extend beyond a single incident and are reported alongside debt problems and broken family relationships. In 68.0\% of Reddit posts recording reactions, older adults showed resistance or continued involvement; in Baidu Tieba family-confrontation posts, 41.9\% showed continued belief. Relatives identified fraud in 74.4\% of Reddit posts and narrated cases on Baidu Tieba, but without the victim's consent, relatives may lack the legal authority to intervene. In Reddit posts recording reactions, account locks or credential changes appeared in 8.9\% under resistance or continued involvement, compared with 33.3\% under acceptance. Measures such as phone or network monitoring and financial takeover were reported more often under resistance. Recovering funds is very limited in posts recording follow-up results of fraud; Baidu Tieba users reported that police classified older adults' voluntary transfers as civil disputes. 
\end{abstract}

\input{introduction}

\input{related_work}

\input{Methods}

\input{Results}

\input{Limitations}
\input{Summary}

\bibliographystyle{IEEEtran}
\bibliography{references}
\input{Appendix}

\end{document}

%% file: introduction.tex
\section{Introduction}

In a Reddit post, a family member wrote about a grandfather who ignored the police warning about the ongoing fraud scheme and, citing his bank's alerts on small withdrawals, said, ``I don't think it's a scam.'' Although the family noticed the abnormality and sought help, the grandfather trusted his understanding of the bank's alert policy and did not follow his family's and law enforcement's advice, delaying intervention. Only later could the family hold a family meeting and, together with an aunt and a relative holding a power of attorney, schedule a bank meeting. This case exposes a core contradiction in the elder fraud response: discovering fraud, making older adults aware of the threat, and having legal authority to intervene rarely happen simultaneously.

Financial security in later life is essential to households as the global population aged 60 and above is expected to increase from 1 billion in 2019 to 1.4 billion by 2030~\cite{WHO-Ageing-2030}. Older adults have diverse cybersecurity needs and face elder scams~\cite{GoodIT23-OlderAdults}, and in 2024, complainants aged 60 and above reported \$4.9 billion in losses to the Internet Crime Complaint Center~\cite{IC3-24}. In a December 2022--February 2023 survey of 4,293 Chinese internet users aged 50 and above, 129 (3.00\%) reported online fraud loss~\cite{Wang25-OnlinePerils}. By the end of 2022, Chinese police reported solving over 39,000 elder-fraud cases and recovering over RMB 30 billion in stolen proceeds and losses~\cite{MPS-22}.
In the Federal Trade Commission's 2022--2023 report, over 16\% of reports for adults aged 80 and above were submitted by third parties~\cite{FTC-23}.

These aggregate statistics say little about how scams unfold or what relatives do when they notice them. Official channels and online communities document suspected scams, yet families struggle to stop an active scam. For example, relatives ask forums how to spot scams and support victims~\cite{SOUPS25-Oak,CCS25-Advice,CHI25-EmotionHelp}, yet in these circumstances, older adults often deny the scam, and relatives trying to prevent losses may lack authority to intervene without the account holder's consent. 
Still, prior work has not associated detection and continued belief with family's protective actions (relatives' account locks or credential changes, reports or legal action, financial takeover, phone or network monitoring, or physical intervention) and remedies across platforms and languages. The closest prior forum studies each examine one platform~\cite{SOUPS25-Oak,CCS25-Advice,Lee21-TiebaFraud,CHI25-FamilySafeguard}. To the best of our knowledge, no prior work has examined online discussions of elder scams across two platforms in two languages; we compare English discussions on Reddit with Chinese discussions on Baidu Tieba.

In this paper, we investigate Reddit and Baidu Tieba discussions on how families detect, confront, and respond to elder scams. Prior forum studies described these steps qualitatively~\cite{CCS25-Advice,CHI25-FamilySafeguard}; in our multi-year elder-scam corpora, we find that on both platforms older adults often resist warnings and that few posts report recovering lost money. 
We investigate four research questions: 

\textbf{RQ1:}~How do reported financial and non-financial costs differ between single-incident and repeated elder scams? 

\textbf{RQ2:}~Which scam categories and persuasion tactics relate to continued belief in scams among older adults? 

\textbf{RQ3:}~Who detects suspected elder scams or narrates first-hand personal cases in online discussions? 

\textbf{RQ4:}~What protective actions, remedy channels, and recovery outcomes do households report in elder scams?

To answer these research questions, we collected Reddit posts across nine subreddits from May 2021 to June 2026 via Arctic Shift~\cite{ArcticShift} and Baidu Tieba posts from 2007 to 2026 by crawling 157 public forums with a custom crawler. 
We then applied keyword filtering, retaining 118,157 Reddit and 1,341,850 Baidu Tieba thread-opening posts. Subsequently, we used a large language model (Qwen3.7-Flash~\cite{Qwen37Flash} for Reddit and GPT-5.5~\cite{OpenAI-GPT55-SystemCard} for Baidu Tieba) to identify posts relevant to elder scams. 
As a result, we obtained a novel and unique dataset of 13,082 Reddit posts and 3,933 Baidu Tieba posts. To answer RQ1, we compared reported losses and harms between single-incident and repeated elder scams. To address RQ2, we analyzed Reddit stances by category and tactic alongside continued belief in Baidu Tieba confrontation posts and across all posts in each Tieba category. To investigate RQ3, we quantified recorded detectors on Reddit and qualitatively examined relatives' accounts on Baidu Tieba. To probe RQ4, we compared families' protective actions across stances and evaluated remedy channels and recovery across institutions and assets.

The contributions of this paper are: (1) We provide a novel and unique dataset of Reddit and Baidu Tieba discussions capturing family accounts of elder scams, as well as present aligned codebooks for both platforms. (2) Analyzing reported losses, borrowing, and harms, we found repeated or ongoing victimization in 48.7\% (5,566 out of 11,428) of the Reddit posts that state whether the scam happened once, repeatedly, or was still ongoing (1,654 posts did not and were excluded). On Baidu Tieba, repeated extraction reached 69.8\% (2,619 out of 3,752 posts coded One-Off or Repeated; 181 N/A posts excluded). Repeated cases on Baidu Tieba involve higher stated losses than single incidents, while repeated cases on Reddit report damaged family relationships more often. (3) In response to warnings, older adults showed resistance or continued involvement in 68.0\% (2,705 out of 3,979 stance-coded Reddit posts). 
On Baidu Tieba, 41.9\% (443 out of 1,057 confrontation posts) record continued belief among older adults. Reddit resistance or continued involvement is more common in romance and investment than in phishing or authority impersonation. In Tieba, across all posts in each category, including posts without confrontation outcomes, continued belief concentrates in multi-level marketing, investment, and health-product sales. (4) Investigating discovery, we found relatives are the primary recorded detectors on Reddit at 74.4\%; Baidu Tieba accounts also describe relatives' efforts to identify suspected scams. (5) Across stances, account locks or credential changes appear in 33.3\% of Reddit posts where the older adult accepted warnings. However, under resistance or continued involvement, this share is 8.9\%. Phone or network monitoring and financial takeover appear more often under resistance or continued involvement. Out of the posts that state a recovery outcome, full or partial recovery is 5.5\%  on Reddit and 9.4\% on Baidu Tieba. 

Overall, our work provides a first-hand overview of elder scams and barriers to intervention. Accounts describe relatives recognizing suspected fraud while struggling to obtain cooperation or institutional assistance, so despite third-party reporting and trusted-contact rules~\cite{FTC-23,FINRA-SeniorRules}, clearer remedy channels are needed. We hope this study's findings can facilitate protective holds subject to review and autonomy safeguards, contributing to elder financial security.

%% file: related_work.tex
\section{Related Work}

\textbf{Scam Mechanisms and Help-Seeking:} Prior work has analyzed how attackers reach targets and induce payments in technical support, SMS phishing, and cryptocurrency scams~\cite{Miramirkhani17,CCS26-Scambaiters,SP24-SMSPhishing,SP24-CryptoConman,NDSS25-CtrlAltDeceive,NDSS26-LOKI}. Lyu et al.~\cite{CCS26-ScammersTalkBack} found that malicious callers used persuasion to establish authority and rapport while they escalated requests for personal information. 
Bouma-Sims et al. examined teenagers' and adults' judgments of YouTube giveaway scams~\cite{NDSS25-Kids}. Interview studies of pig-butchering scams described trust building and victim exploitation~\cite{SP24-PigButcher}. On Reddit, Oak and Shafiq~\cite{SOUPS25-Oak} characterized fraud types and community support. Bouma-Sims et al.~\cite{CCS25-Advice} analyzed advice and support in Reddit fraud discussions.
Chanenson et al.~\cite{CHI25-EmotionHelp} linked victims' emotional responses and help requests to fraud stages. Other studies measured self-reported financial losses in Chinese and European surveys~\cite{Lei22-CHARLSFraud,Kemp23-EUFraudImpact} and in crypto-scam complaints and UK victim interviews~\cite{Ordekian24-PigButchering,Havers24-Reporting}. 

\textbf{Family Safeguarding in Elder Fraud Victimization:} Previous studies have documented fraud victimization among older adults~\cite{Burnes17,DeLiema20-Fraud}. Parti and Tahir~\cite{Parti23-Underreporting} interviewed older Americans and found that embarrassment and shame hindered fraud reporting. DeLiema~\cite{DeLiema17} distinguished third-party fraud from financial exploitation by trusted individuals, and Cross~\cite{Cross19} compared sustained interpersonal pressure in romance fraud with psychological abuse in domestic violence. Deng et al.~\cite{CHI25-FamilySafeguard} analyzed RedNote elder fraud posts and comments and identified five protective functions. Sharevski and Vander Loop~\cite{sharevski2026scams} interviewed sighted families and families with blind or low vision members about AI voice cloning scams.
Other work has also examined household security, proxy access, and tech caregiving~\cite{Murthy21-Households,Latulipe22-Proxies,Havens24-ProxyAccounts,Kropczynski21-TechCaregiving}. 
Similarly, research on refugee populations and assistive technologies suggests that security design must consider both helpers' access risks and users' actual needs~\cite{SP18-Refugees,SP23-Cybercafes,SP22-AtRiskSoK}.

\textbf{Analyzing Online Discussions of Fraud:} 
Lee~\cite{Lee21-TiebaFraud,lee2022online} analyzed Chinese Baidu Tieba fraud narratives and crime scripts. Vafa et al.~\cite{Vafa2025a} found that several popular Telegram VPN channels shared malicious files presented as free VPN services. Olteanu et al.~\cite{Olteanu19-SocialData} showed that platform structures, collection time spans, and discussion formats determine what social media datasets capture. 

While prior works examine these dimensions separately, our work examines Reddit detectors and relatives' accounts in first-hand personal-case Tieba threads alongside older adults' responses to warnings (including resistance and continued belief), family protective actions, institutional responses, and stated recovery outcomes.

%% file: Methods.tex
\section{Methodology}\label{sec:corpus-annotation}

\begin{figure*}[!h]
\centering
\includegraphics[width=0.9\textwidth]{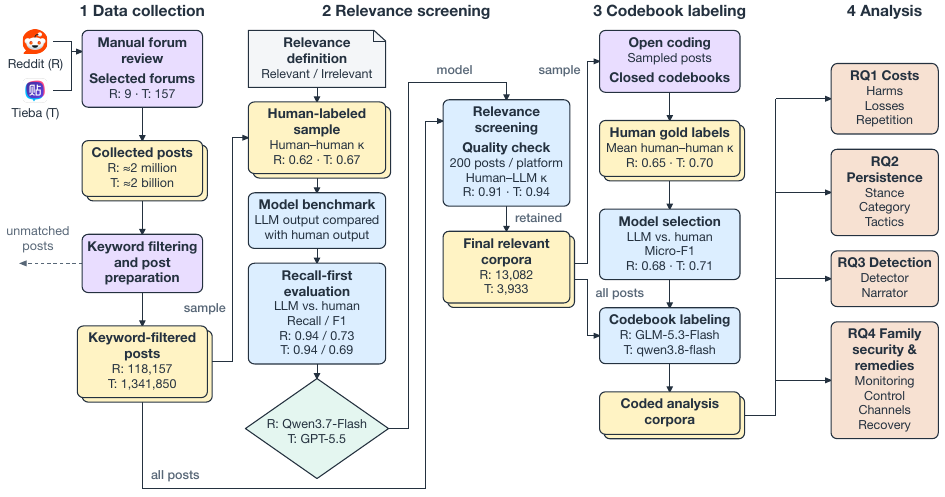}
\caption{Workflow for data collection, screening, coding, and analysis. }
\label{fig:tieba-workflow}
\end{figure*}

We illustrate the four steps to collect and label our datasets on Reddit and Baidu Tieba in Figure~\ref{fig:tieba-workflow}. On both Reddit and Baidu Tieba, we (1) obtained publicly available posts and removed posts that were either deleted or moderated, (2) filtered posts based on keywords specific to each platform, (3) hand-labeled samples to detect false positives, tested machine learning models and large language models (LLMs), and checked keyword-filtered posts for actual elder scam cases, and (4) developed codebooks on double-coded samples, tested models against human labels, and labeled our final datasets.

\subsection{Data Collection}

\textbf{Reddit}: We searched for Reddit communities that discuss elder scams. We searched 309 subreddits in the Arctic Shift archive~\cite{ArcticShift} and read sample posts (927). Since keyword queries on Arctic Shift repeatedly timed out, we did not collect a keyword sample directly from the archive. Instead, we downloaded raw posts from Arctic Shift first and ran keyword filtering.
We selected 9 primarily English-language subreddits: r/Scams, r/legaladvice, r/personalfinance, r/ScamNumbers, r/AgingParents, r/Romancescam, r/CaregiverSupport, r/eldercare, r/FraudPrevention. We retrieved titles, body, date, score, and comment counts for roughly 2 million posts (May 2021 to June 2026) from those communities. We removed \textit{deleted} or \textit{moderated} posts (30.2\%) before keyword filtering. This left us with 1,576,977 usable posts.
Since all 9 subreddits are primarily English-language, we did not filter by language in data collection. We checked 50K random posts in the keyword-filtered pool with fastText~\cite{Joulin2017BagTricks} and found that 49,827 titles and bodies were in English (99.65\%). 

\textbf{Baidu Tieba}: Since elder-scam posts on Baidu Tieba are scattered across general forums, we searched in Chinese. We used phrases like ``an older adult was scammed", ``an older adult in my family was scammed", ``what to do when an older adult is scammed" (full list of queries is in this anonymous repository \footnote{https://anonymous.4open.science/r/elder-scam-artifacts-7B5F \label{fn:shared}}).
After reading samples in 766 forums, we selected 157 forums in June 2026. We retrieved roughly 2 billion public posts. 
Our script used a dynamic buffer where it automatically filters out deleted or removed posts. 
We kept all dated posts from 2007 to 2026 because a 5-year window would have too few posts in Baidu Tieba for category-level comparisons. This 20-year window covers major shifts in education and internet use among older adults in China~\cite{NBS-Census7-QA21,CNNIC-23,CNNIC-57}. 

\noindent\textbf{Keyword Filtering:} Online forums have both scam reports and regular questions. To extract candidate posts, we used English and Chinese keyword rules (full lists of keywords and rules are in the anonymous repository \footref{fn:shared}). On Reddit, we filtered with direct terms (e.g., ``elder fraud", ``senior scam", ``elder financial abuse"), family member terms, and specific scam types in titles and bodies to keep 118,157 unique posts. On Baidu Tieba, we filtered using Chinese elder-scam phrases and word-pairing rules in titles and available opening text to keep 1,341,850 unique posts. Many general words like ``insurance", ``health products", ``investments", and ``collectibles" are common in regular questions. If a post only matched those general words, we only kept it if it also contained both an older adult or relative and fraud, loss of money, payment, complaint, or manipulation (e.g., ``older adult", ``insurance", ``was scammed" together, instead of just ``insurance").

\noindent\textbf{Relevance Screening}: Keyword filtering could still have false positive posts. To investigate this, we created another pipeline in which two coders iteratively coded 12,353 randomly sampled Reddit posts and identified 1,697 true-positive elder-scam posts (Cohen's $\kappa$ = 0.62). On the other hand, of the 17,002 randomly selected Baidu Tieba posts, 1,549 were true positives (Cohen's $\kappa$ = 0.67); We used these two true-positive sets as the ground truth for our LLM labeling.

Based on these reference posts, we set aside 5,000 Reddit posts (687 Relevant, 4,313 Irrelevant) and 5,000 Baidu Tieba posts (456 Relevant, 4,544 Irrelevant) stratified by label and forum as evaluation sets. 
Local models were trained on the remaining 7,353 Reddit posts and 11,984 Baidu Tieba posts after excluding text duplicates of evaluation posts. On the test sets of 5,000 posts, we compared traditional machine learning models, fine-tuned language models (e.g., BERT, RoBERTa), and pretrained LLMs (given our screening rules directly without fine-tuning) (Tables~\ref{tab:bench-tieba-tp} and~\ref{tab:bench-reddit-tp} in Appendix~\ref{app:benchmarks}). Like previous studies on scam forums~\cite{CHI25-EmotionHelp}, we prioritized high recall to retain borderline elder scam cases, accepting lower precision. Fine-tuned models had higher overall $F_1$ scores (0.76 to 0.80 on Reddit; 0.70 to 0.77 on Baidu Tieba). But they missed more real cases with lower recall (0.73 to 0.91; 0.72 to 0.85). On Reddit, Qwen3.7-Flash~\cite{Qwen37Flash} (medium reasoning configuration, temp = 0) identified 93.7\% of the relevant posts (Recall = .937). Its precision was .591, its $F_1$ score was .725, and its Cohen's $\kappa$ was .670 compared to human labels. On Baidu Tieba, the best-performing model was gpt-5.5 (xh reasoning configuration, temp = 0), which achieved a recall of 0.936, a precision of 0.5524, and an $F_1$ score of 0.694.

We applied our high-recall models to all keyword-filtered posts. On Reddit, we used Qwen3.7-Flash (medium reasoning configuration) to screen all 118,157 keyword-filtered posts. We retained 13,082 flagged posts 
to create our final Reddit dataset.
On Baidu Tieba, we used GPT-5.5 with extra-high reasoning effort~\cite{OpenAI-GPT55-SystemCard} to screen all 1,341,850 keyword-filtered posts. We retained 3,933 candidate posts. 
To verify screening accuracy, we randomly sampled 200 posts from each platform, and two coders independently annotated them for relevance, achieving a Cohen's $\kappa$ of 0.91 on Reddit and 0.94 on Baidu Tieba against the LLM labels.

\subsection{Codebook Development and Groudtruth}

We developed two codebooks for Reddit and Baidu Tieba. We focused on three stages of scams: (1) scam mechanics, (2) detection and response, (3) aftermath and recovery (Figure~\ref{fig:codebook-overview}). The complete codebooks are in Appendix~\ref{app:codebooks}. 
In Figure~\ref{fig:codebook-overview}, orange nodes denote high-level lifecycle dimensions and sub-dimensions, while leaf nodes are color-coded by platform coverage: white boxes indicate sub-themes shared across both codebooks, blue boxes indicate Reddit-only sub-themes, and yellow boxes indicate Baidu Tieba-only sub-themes.

\begin{figure}[t]
\centering
\includegraphics[width=\linewidth]{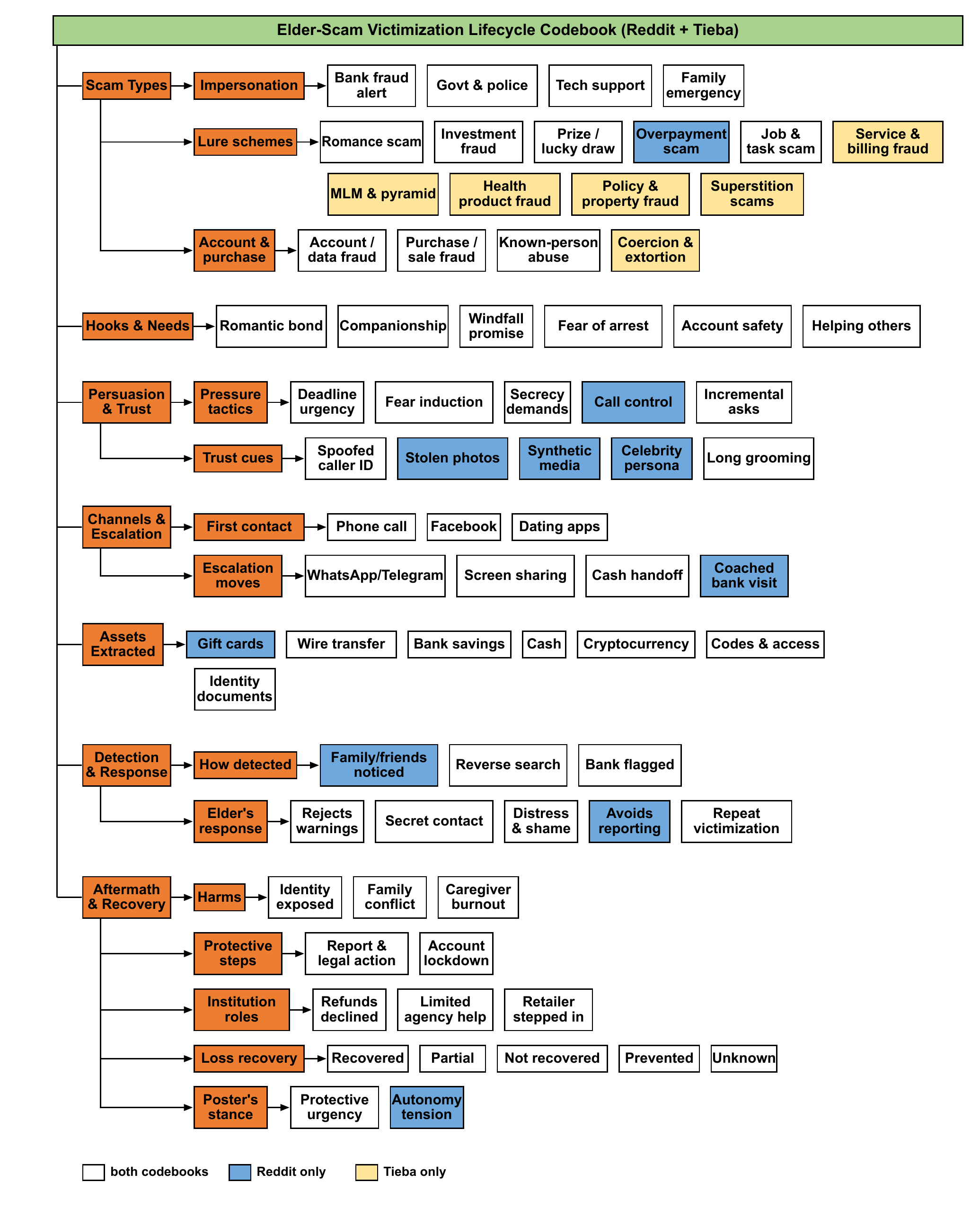}
\caption{Elder Abuse Scams hierarchical codebook.}
\label{fig:codebook-overview}
\end{figure}

On Reddit, one author hand-labeled 200 sampled posts and used an open coding process~\cite{glaser2017discovery} to determine themes and sub-themes to create a 16-field codebook, comprising four single-select categorical fields (e.g., repetition pattern, loss recovery status), one bounded field allowing up to two labels (scam category), and 11 unconstrained multi-label fields (Appendix~\ref{code-reddit}). On Baidu Tieba, one author hand-labeled 227 sampled posts and used an open coding process to create a 12-field codebook, comprising four single-select categorical dimensions and eight bounded multi-label dimensions, alongside one auxiliary binary field indicating whether the poster was a first-hand participant (Appendix~\ref{code-tieba}). 

We had two coders independently label all 200 Reddit posts and 227 Baidu Tieba posts using these codebooks. The average Cohen’s $\kappa$ was 0.65 for Reddit and 0.70 for Baidu Tieba, indicating \textit{substantial agreement}. We then used these two sets of double-coded posts as the gold standard to test our model's labeling on each platform. 
Reddit and Baidu Tieba have different forum ecosystems and scam distributions. So, the two codebooks track corresponding themes but keep platform-specific definitions instead of treating counts identically. For example, on Reddit we track who detected the scam, whereas on Baidu Tieba we examine relatives' accounts in first-hand personal-case posts.

\textbf{Multi-Field Labeling of the Final Corpora}: 
To select production models for multi-field coding, we partitioned our double-coded reference sets (200 Reddit and 227 Baidu Tieba posts) into in-context prompt examples and development evaluation splits. Specifically, we selected 39 Reddit posts and 26 Baidu Tieba posts illustrating positive and boundary cases across codebook dimensions as few-shot prompt examples, reserving the remaining 161 Reddit and 201 Baidu Tieba posts for model benchmarking (Tables~\ref{tab:reddit-coding-full} and~\ref{tab:tieba-coding-full} in Appendix~\ref{app:benchmarks}). Our best models achieved test micro-$F_1$ scores of 0.68 on Reddit (Table~\ref{tab:reddit-coding-full}) and 0.71 on Baidu Tieba (Table~\ref{tab:tieba-coding-full}). Accordingly, we used GLM-5.3-Flash~\cite{GLM53Flash} to label all 13,082 Reddit posts and \texttt{qwen3.8-flash} to label all 3,933 Baidu Tieba posts. We also tested some lexical- and embedding-based models, but all of them were outperformed by the LLMs. 

\textbf{Final Quality checks on the corpus}: To verify that the LLM outputs matched human annotations, we re-annotated random 200-post samples from each corpus, obtaining average Cohen's $\kappa$ scores of 0.62 on Reddit (16 fields) and 0.65 on Baidu Tieba (12 variables). We then compared these newly labeled 200 sample sets with their corresponding LLM labels and obtained $F_1$ scores of 0.73 for Reddit and 0.71 for Tieba. Because our earlier labels for Elder’s Response to Warning and Informal Family Security Work (column 9 and 10 for Reddit codebook Appendix ~\ref{app:codebooks}) had low $F_1$ (0.54) scores, we asked two coders to iteratively relabel a fresh random sample of 1,000 relevant posts and achieved a Cohen’s Kappa of 0.72 and an $F_1$ of 0.89 compared to the LLM label. The resulting associations matched those on the full LLM-labeled corpus in Section~\ref{sec:res-security-work} (e.g., higher account-lockdown rates under warning acceptance than under resistance: 21.6\% vs.\ 7.8\%, compared to 33.3\% vs.\ 8.9\% in the full corpus), confirming that our downstream analyses remain robust to labeling noise.

%% file: Results.tex
\section{RQ1: The Cost to Families}
\label{sec:res-cost}

Based on our codebooks, we characterized older-adult scams across Reddit and Tieba. The 13,082 Reddit posts and 3,933 Tieba posts spanned digital and commercial schemes. In this section, we first describe scam categories and tactics as background for our RQs (Section~\ref{sec:res-landscape}) and then compare resulting harms across single-incident and repeated scams (Section~\ref{sec:res-harms}). Note that we quote Reddit posts in short verbatim snippets, whereas we translate Tieba examples (see Ethics Considerations~\ref{ethics}). 

\subsection{Scam Categories, Contact Channels, and Persuasion Tactics}
\label{sec:res-landscape}

\begin{figure}[t]
  \centering
  \subfloat[Reddit\label{fig:reddit-categories}]{%
    \includegraphics[width=.49\columnwidth]{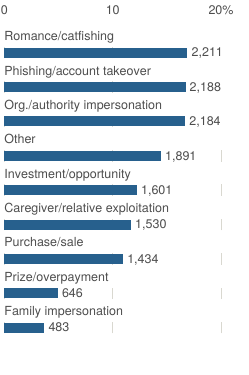}}\hfill
  \subfloat[Tieba\label{fig:tbcats}]{%
    \includegraphics[width=.49\columnwidth]{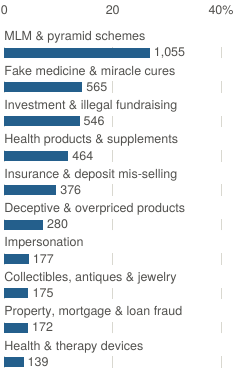}}
  \caption{Overlapping fraud categories (entire-corpus denominators): (\protect\ref{fig:reddit-categories}) all nine on Reddit ($N=13{,}082$) and (\protect\ref{fig:tbcats}) top-ten on Tieba ($N=3{,}933$). Bars represent percentages, labels post counts.}
  \label{fig:category-comparison}
\end{figure}

\begin{figure}[t]
  \centering
  \subfloat[Tieba ($n=897$).\label{fig:tbtrends}]{%
    \includegraphics[width=1\columnwidth]{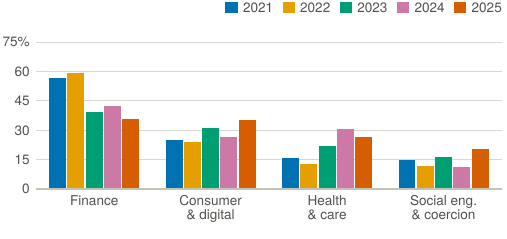}}\\[6pt]
  \subfloat[Reddit ($n=11{,}979$).\label{fig:rdtrends}]{%
    \includegraphics[width=1\columnwidth]{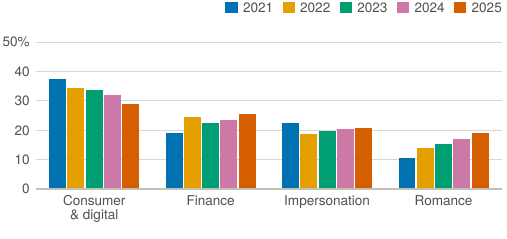}}
  \caption{Annual shares of scam category clusters from 2021 to 2025 on Baidu Tieba and Reddit}
  \label{fig:category-trends}
\end{figure} 

\textbf{Scam categories and targeted assets:} Figure~\ref{fig:category-comparison} shows our Reddit and Tieba scam categories, whose shares sum to over 100\%. Reddit users mainly reported remote digital scams, with similar shares of romance scams, phishing/account takeover, and authority impersonation. However, Tieba posts focused on offline-to-online commercial schemes, including multi-level marketing (MLM), health-product sales, and deceptive and overpriced product sales.
Figure~\ref{fig:tbtrends} shows Tieba category-cluster shares from 2021 to 2025 across 897 of 3,889 dated posts. As shares of each year's posts, the Finance cluster fell from 57\% to 36\%, while the Health and care cluster rose from 16\% to 27\%. Figure~\ref{fig:rdtrends} shows the same shares for four Reddit clusters, which reuse two of Figure~\ref{fig:tbtrends}'s names but group Reddit's own categories, across 11,979 of 13,082 posts: Consumer and digital (phishing or account takeover, purchase or sale, prize or overpayment), Finance (investment and caregiver or relative exploitation), Impersonation (of organizations, authorities, or relatives), and Romance. Consumer and digital fell from 38\% to 29\%; Finance rose from 19\% to 26\% as investment scams rose from 6\% to 15\%; and Romance rose from 10\% to 19\%. These changes hold within r/Scams, which grew from 47\% to 63\% of each year's posts.
For targeted assets, Reddit posters named cash ($n= 1{,}950$), bank accounts ($n= 1{,}292$), personal information only ($n= 1{,}057$), gift cards ($n= 602$), cryptocurrency ($n= 551$), wire transfers ($n= 529$), and ``other'' ($n= 2{,}267$). Interestingly, 34.0\% of Reddit posts (4,444 posts) described attempts without financial loss.
On Tieba, posters reported monetary transfers ($n= 3{,}229$; coded as Money in Table~\ref{tab:asset-recovery}), health damage ($n= 794$), and contracts or debt ($n= 643$).

\begin{figure}[t]
  \centering
  \includegraphics[width=.96\columnwidth]{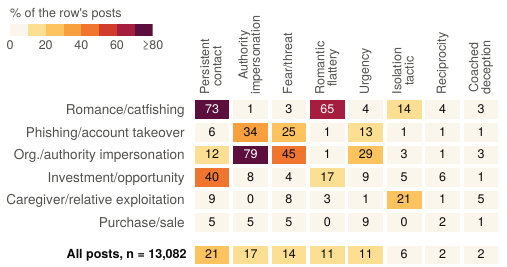}
  \caption{Persuasion tactics ($\geq150$ posts): all-post percentages in Reddit's six largest categories excluding Other (Figure~\ref{fig:reddit-categories}).}
  \label{fig:reddit-tactics}
\end{figure}

\textbf{Reddit persuasion tactics and trust cues:} The divide between coercion and grooming is shown in Figure~\ref{fig:reddit-tactics}.
Posts about authority impersonation described coercion via institutional mimicry as a trust cue (81\%), the authority-impersonation tactic (79\%), and fear or threats (45\%). Phishing posts described institutional mimicry as a trust cue less often (48\%), and family-impersonation posts ($n= 483$) described urgency (53\%) and fear or threats (40\%). In an authority-impersonation case, one poster wrote that callers ``[t]old her she would be charged with international money laundering and fined \$10K if she didn't immediately approve the withdrawal, started threatening her and confusing her''. Conversely, posters reported grooming in romance and ``catfishing'' scams via ``building relationships first and then making demands'' (77\%), daily contact (73\%), flattery (65\%), and photos or videos (43\%). In a romance case, one poster noted that scammers ``continue to have a new story, medical emergency, Lottery winnings, you name it excuse for him to send more money to her to keep him hooked''. Posts about caregiver and relative exploitation described the isolation tactic (21\%) more than all other categories. In one caregiver case involving urgent payment demands, a poster described a caller who ``needs the money within 5-10 minutes or he sends my GG 10+ messages and spam calls her within 3 minutes''. Across Reddit, we found four overlapping trust cues: (a) institutional mimicry (25\%), (b) relationship building before the ask (18\%), (c) real or stolen photos or videos (10\%), and (d) real personal or account information (7\%). Across the six largest categories excluding Other, this last cue reached 20\% in authority impersonation and 10\% in phishing. Conversely, institutional mimicry led in purchase scams (29\%), and relationship building before the ask led in investment (32\%) and caregiver and relative exploitation (23\%).

\begin{figure}[!ht]\centering\includegraphics[width=0.83\columnwidth]{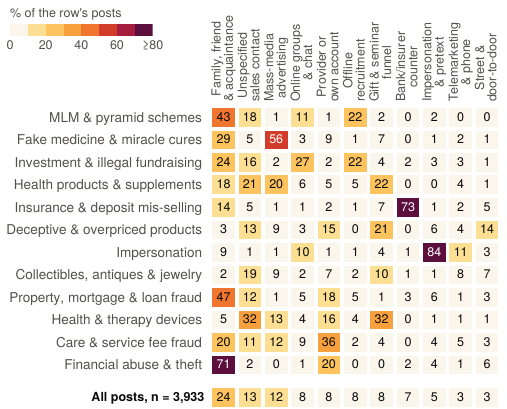}
\caption{Overlapping contact-channel percentages ($\geq120$ posts) for all posts in Tieba's 12 largest categories.}
\label{fig:tbchannel}\end{figure}

\begin{figure}[t]\centering\includegraphics[width=0.82\columnwidth]{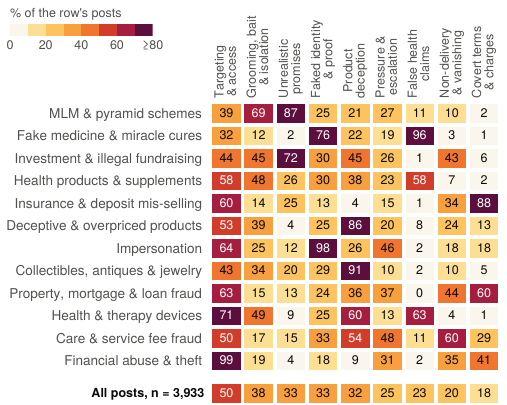}
\caption{Nine overlapping tactic families: percentages of all category posts within Tieba's 12 largest categories.}
\label{fig:tbtactic}\end{figure}

\textbf{Tieba contact channels and persuasion tactics:} To examine how scams in our Tieba dataset reached older adults, we analyzed the 12 largest scam categories by contact channel and persuasion tactic. Figure~\ref{fig:tbchannel} shows that reported contact channels extended beyond stranger cold calls. Often, scams entered through familiar social ties or offline institutional venues. We found family, friends, and acquaintances served as entry channels in 43\% of MLM and 71\% of financial-abuse posts. Bank and insurance counters accounted for 73\% of insurance and deposit mis-selling posts, and mass media appeared in 56\% of fake-medicine posts, showing offline settings remained key exposure points for older adults. Figure~\ref{fig:tbtactic} illustrates persuasion tactics (measured via the nine red-flag codes) across these 12 Tieba categories. We found that 87\% of MLM posts described high-return and recruitment promises, while 96\% of fake-medicine posts described fabricated claims of healing or health benefits. For instance, posters recounted bank-counter staff steering older savers into multi-year insurance policies disguised as regular deposits, and community ``free-gift'' health lectures invoking fabricated medical authority to peddle costly supplements or treatment devices.  

\subsection{Financial Losses, Repeated Extraction, and Non-Financial Harms}\label{sec:res-harms}

\begin{figure*}[!t]\centering\includegraphics[width=.86\textwidth]{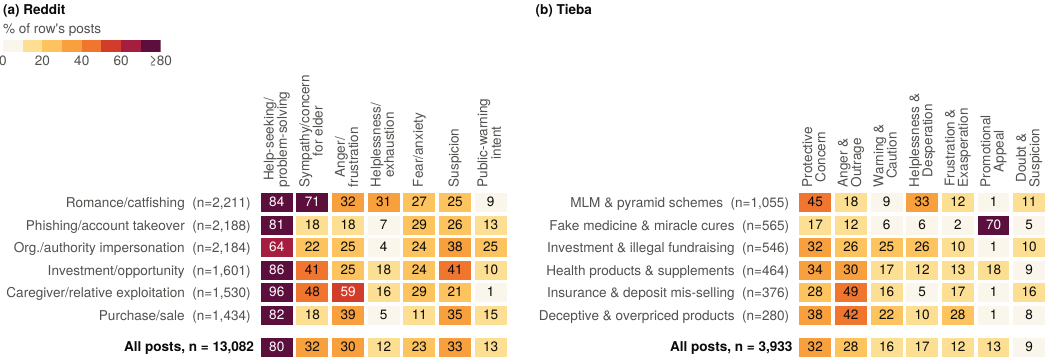}
\caption{Top seven multi-label emotions in each platform's six largest categories excluding Other, as row post percentages}
\label{fig:emotcat}\end{figure*}

\begin{figure}[!htb]\centering\includegraphics[width=.94\columnwidth]{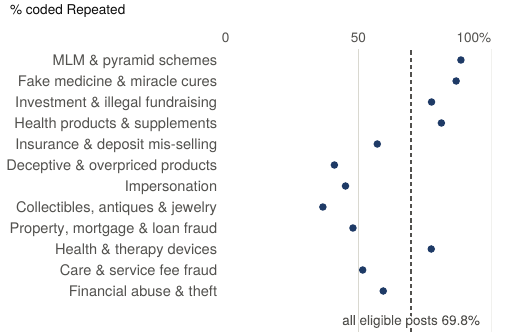}
\caption{Repeated extraction in Tieba's 12 largest overlapping categories ($\geq120$ posts). Fractions show Repeated over Repeated-or-One-Off posts; the dashed line marks the 69.8\% corpus rate ($n=2{,}619$ of $n=3{,}752$; $n=181$ N/A excluded).}
\label{fig:tbrepeated}\end{figure}

\begin{figure*}[!ht]\centering\includegraphics[width=.8\textwidth]{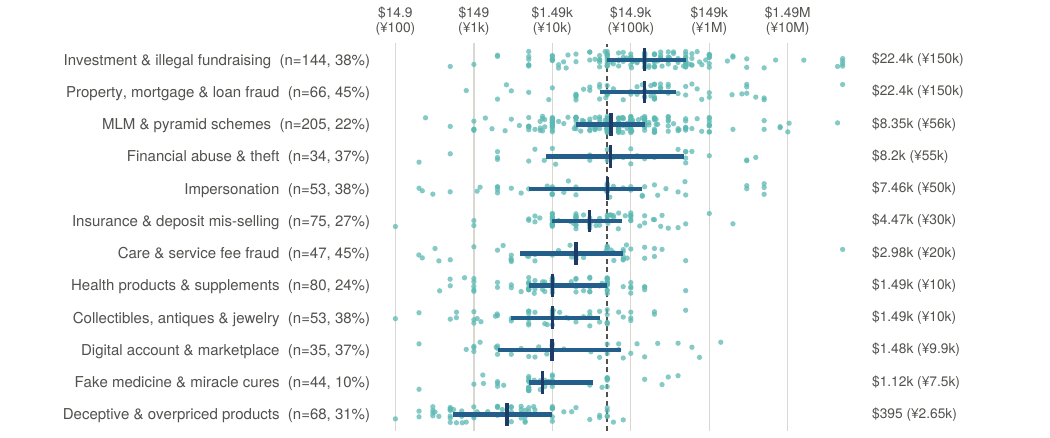}
\caption{Stated Tieba losses in 12 categories ($\geq30$ first-hand loss reports), log-scale USD (RMB, US\$1=RMB~6.70455, ECB, October~1, 2026). Dots show posts, bars IQR, ticks medians, and the dashed line the $n=804$ median (aggregate-only reports excluded; values below the displayed range not shown). }
\label{fig:tbloss}\end{figure*}

\textbf{Stated financial losses and debt burden:} We extracted stated losses on Reddit in 994 posts (7.6\%), finding a median of \$8,000 (IQR \$1,000--\$45,000) and a 90th percentile of \$200,000. Keyword-window extraction on 3,094 first-hand Tieba posts retrieved amounts in 804 posts (26.0\%), with a median of RMB~49,800 (about \$7,400).  
Figure~\ref{fig:tbloss} illustrates median losses across the 12 Tieba categories with at least 30 first-hand loss reports, from about RMB~2,650 for deceptive and overpriced product sales to about RMB~150,000 for both investment and illegal fundraising and property, mortgage, and loan fraud. Tieba posts also described borrowing money to pay scammers. When relatives restricted cash for a cure-all course, one Tieba poster wrote: ``She did not have enough money, so she borrowed it. Different creditors came to our home six or seven times. \ldots{} The conflict within the family is very intense. \ldots{} Everyone in the family is at a loss. What should we do?''

\begin{figure}[!t]
  \centering
  \subfloat[Ongoing or repeated scam ($n=11{,}428$ coded; $n=1{,}654$ N/A excluded).\label{fig:reddit-repeat}]{\includegraphics[width=.5\linewidth]{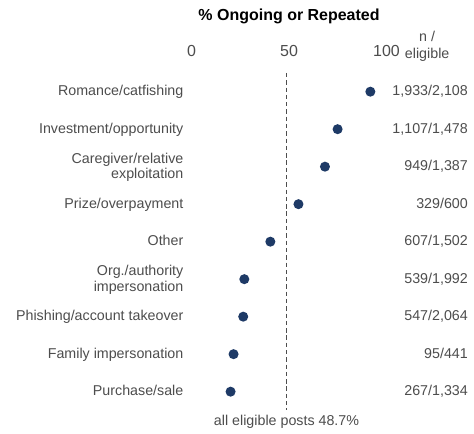}}\hfill
  \subfloat[Older adult denied fraud or defended scammer ($n=3{,}979$ coded; $n=9{,}103$ unrecorded excluded).\label{fig:reddit-denial}]{\includegraphics[width=.5\linewidth]{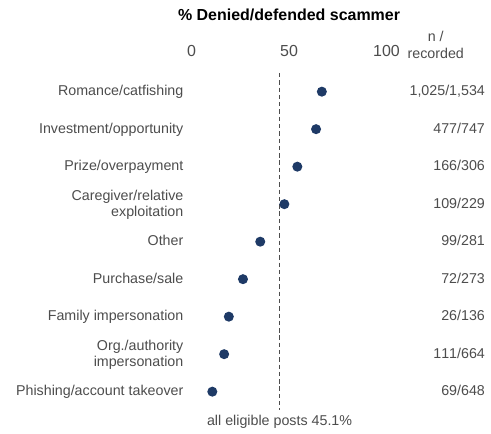}}
  \caption{Reddit repetition and fraud denial or scammer defense across overlapping categories ($N=13{,}082$). Fractions show within-category counts over category-specific coded denominators; dashed lines mark overall panel rates (48.7\% and 45.1\%).}
  \label{fig:reddit-repeat-denial}
\end{figure}

\textbf{Repeated or ongoing victimization and extraction:} On Reddit, posters reported fraud still ongoing at posting in 4,118 posts (31.5\%). Figure~\ref{fig:reddit-repeat} shows that among 11,428 posts with a stated repetition pattern (1,654 N/A excluded), repeated or ongoing victimization appeared in 5,566 posts (48.7\%) across categories, reaching 91.7\% (1,933 of 2,108) in romance scams. A poster mentioned their 72-year-old father was ``usually talking to multiple scammers at the same time'' and had engaged ``50+ different scammers'' over 6 years. They added that he ``still sends gift cards''. Similarly, Figure~\ref{fig:tbrepeated} shows repeated extraction in 2,619 of 3,752 coded Tieba posts (69.8\%) via repeated purchases, phased fees, or recurring payments, exceeding 75\% in MLM, fake-medicine, and health-product scams. Conversely, repeated extraction in deceptive and overpriced product sales and collectibles stayed below 50\%. Across 804 loss posts (1 lacking repetition coding), median losses were about RMB~50,000 in 510 repeated-extraction posts (IQR RMB~10,000--200,000) versus about RMB~30,000 in 293 single-incident posts (IQR RMB~5,000--100,000). In investment and illegal fundraising, medians reached about RMB~150,000 in both groups (103 repeated-extraction and 41 single-incident posts), indicating high-ticket schemes could cause severe loss in a single incident. In one Tieba case, a daughter recounted that a caller demanded fees ``again and again'', starting with RMB~3,000 and RMB~3,500 before ``swindling 51,000 in all''. Relatives could also bear some of the financial cost in these scams. One adult child on Tieba reported that their mother had lost tens of thousands of Yuan in an investment scheme. When she later incurred gambling debts, the adult child helped repay more
than RMB~100,000.

\textbf{Non-financial harms and emotions in help-seeking posts:} Reddit posters also reported emotional distress or isolation ($n= 2{,}863$), broken family relationships ($n= 1{,}673$), caregiver burnout ($n= 1{,}079$), lost autonomy ($n= 604$), and declining health ($n= 334$). Figure~\ref{fig:emotcat} shows top multi-label emotion codes across categories, led on Reddit by help-seeking/problem-solving ($n= 10{,}401$), with an average of 2.45 emotion codes per post. In contrast, Tieba shows a slightly different ecosystem of emotions. Tieba posters reported worry for the older adult ($n= 1{,}266$), anger and outrage ($n= 1{,}093$), helplessness and desperation ($n= 687$), and warning and caution ($n= 617$), with an average of 1.74 emotion codes per post (at most two per post).

Non-financial harms concentrated in repeated or ongoing victimization, appearing in 45.7\% (2,546 of 5,566) of repeated or ongoing posts versus 17.0\% (995 of 5,862) of single-incident posts. Similarly, broken family relationships appeared in 23.6\% (1,314 posts) of repeated or ongoing posts versus 4.3\% (253 posts) of single-incident posts. By category, non-financial harm reached 71.8\% (1,098 of 1,530) in caregiver and relative exploitation and 53.9\% (1,191 of 2,211) in romance scams. On possible legal intervention, a 72-year-old father's child questioned ``where the line is between: an adult making bad financial decisions vs. someone becoming vulnerable enough that legal intervention may eventually be necessary'' 

\textbf{Summary \& Discussion of RQ1.} Many elder scams in our data occur in stages rather than as one-time thefts. Baidu Tieba posts often involve misleading sales through acquaintances, radio and TV ads, or bank and insurance counters. Previous studies also pointed out that fraud often infiltrates through familiar daily scenarios~\cite{DeLiema17}. Compared to a single incident, repeated cases on Tieba involve higher stated losses, while repeated cases on Reddit report damaged family relationships much more often. Prior research on fraud-support forums also reported similar family stress~\cite{SOUPS25-Oak}. Based on these findings, we provide the following recommendations: (1) \textbf{Banks and payment apps should flag repeated payments over time}, as many scams targeting older adults happen through recurring transfers, fees, or purchases. When transfers, fees, or product charges recur, the app and staff should issue clear warnings to older customers. (2) \textbf{Consumer agencies and senior centers should address offline sales alongside fraudulent calls}, especially promotions in community venues. Senior centers should block deceptive free health lectures, while consumer agencies should monitor misleading counter sales and radio or TV ads targeting older shoppers. (3) \textbf{Family support services should focus on household debt and ongoing family stress}, not just initial loss reporting. Fraud hotlines and family programs~\cite{CHI25-EmotionHelp,CHI25-FamilySafeguard} should help relatives navigate external creditors and family tensions. 

\section{RQ2: Continued Belief and Engagement}
\label{sec:res-persistence}

We analyze how older adults respond to family warnings and when they continue to trust scammers. We also examine which scam categories and persuasion tactics are associated with continued belief and how older adults rationalize the scam.

\textbf{Continued belief and engagement:} We analyzed 3,979 stance-coded Reddit posts (30.4\% of 13,082 posts). Older adults showed resistance or continued involvement (the \textit{Persistence} group) in 2,705 posts (68.0\%). They denied fraud or defended scammers ($n= 1,796$), maintained contact ($n= 805$), expressed anger at relatives ($n= 546$), or downplayed risks ($n= 290$). Conversely, older adults accepted correction in only 1,274 posts (32.0\%). When we excluded posts where only the older adult detected the scam, resistance or continued involvement rose to 76.8\% (out of 3483 posts). On Baidu Tieba, 563 of 3,933 posts (14.3\%) recorded continued belief among older adults (coded \textit{Continue Persist In Belief}), including 443 of 1,057 confrontation posts (41.9\%) coded Dissuade \& Confront Elder. Among confrontation posts, continued belief reached 62.2\% in health and therapy device scams (23 out of 37), 45.7\% in romance scams (16 out of 35), and 29.2\% in collectibles scams (14 out of 48). 

\textbf{Romance and investment scams on Reddit:} Resistance or continued involvement concentrated in scams that build long-term relationships or promise high returns. In stance-coded Reddit posts, 90.7\% (1,391 out of 1,534) of romance-scam posts recorded resistance or continued involvement, and in investment scams, 86.3\% (645 out of 747). However, this share dropped to 28.2\% (183 out of 648) in phishing or account takeover posts, and 31.5\% (209 out of 664) in authority-impersonation posts. We also grouped these posts by persuasion tactic and found resistance or continued involvement in posts using the romantic flattery tactic at 91.2\% (1,028 out of 1,127) versus the authority impersonation tactic at 30.9\% (249 out of 805). One adult child wrote: ``...she got mad and told me even if it's not, it's someone that talks kindly to her, and told me to let her find out on her own.'' In an investment case, a grandchild wrote: ``Grandpa mentioned he had a big increase in his investment already and `did his research to make sure it wasn't a scam'.'' 

\textbf{Continued belief across Tieba categories:} Across all Tieba posts, including posts without confrontation outcomes, older adults showed continued belief in 33.4\% (352 out of 1,055) of posts about MLM schemes and scams. The share in investment and illegal fundraising posts was 21.8\% (119 out of 546), and in health product sales posts 19.4\% (90 out of 464). Conversely, in insurance and deposit mis-selling posts, it was 1.6\% (6 out of 376). One Tieba post shows how a reported payout could sway even a skeptical relative. The poster initially distrusted an investment scheme, but their mother-in-law said she had withdrawn her full RMB~100,000 investment within ten months and still held credits in the scheme. 
In a health-lecture post, a grandchild wrote: ``I have already reported several places like these and recovered more than RMB~100,000. But my grandfather always finds somewhere new to attend lectures. \ldots{} He says, `I believe you that the earlier ones were scammers, but this one definitely is not.'' Another post described relatives spending ``a whole day'' persuading a grandmother to return an oxygen machine until ``one phone call from great-aunt'', the relative who had introduced the seller, ``turned her thinking around''. Overall, recovering money from one merchant may not prevent older adults from falling for new schemes, as scammers anchor their offers in daily social and emotional needs.

\textbf{Explicit denial or defense on Reddit:} Figure~\ref{fig:reddit-denial} shows the distribution of explicit denial or defense: 45.1\% of stance-coded posts involved older adults defending scammers or denying fraud. This proportion varies greatly across fraud types. The share reached 66.8\% in romance and catfishing scams and 63.9\% in investment scams. Conversely, it was 10.6\% for phishing and account takeover and 16.7\% for organization and authority impersonation. 
A poster wrote, ``she replied how am I being scammed? I don't sent [sic] money or gift cards'' after telling an older woman her online partner was a romance scammer. In an investment case, a poster wrote of their father, ``He doesn't believe me. [...] There are people who have made millions he says''. A prize-scam post noted an older woman who ``totally believes these people who call to tell her she's won millions''. These accounts show how older adults rationalized continued belief.

\textbf{Summary \& Discussion of RQ2.} In our data, family warnings often stall when older adults still trust scammers and reject advice. Posts also noted older adults trusting a new scheme even after families recovered a refund. Bouma-Sims et al.~\cite{CCS25-Advice} likewise observed relatives seeking help when confrontation failed to end an older family member's contact with a scammer. Based on these findings, we provide the following recommendations: (1) \textbf{Public anti-fraud education should cover scams that gradually build trust}, not just malicious links or fake caller IDs~\cite{Cross19,SP24-PigButcher}. Research on romance-baiting describes weeks or months of personalized contact, fabricated investment success, and escalating demands for payment~\cite{Gressel26-RomanceBaiting}. Guides for older adults and their families should prioritize concrete actions~\cite{Redmiles2020} and explain warning signs such as daily flattery, fabricated profit evidence, and health lectures pressuring buyers. (2) \textbf{Family support services should help relatives avoid direct family arguments}, as confrontation at home often ends in anger or denial. 
Fraud hotlines and family programs could test support from trusted doctors, old friends, or peers outside the family.  
Adapting clinics where trained technologists help intimate partner violence victims can help provide support for older scam victims. Specifically, continued belief concentrates in long-term relationships such as \textit{romance scams}; thus, effective interventions must address those psychological hooks. 

\section{RQ3: Identifying Suspected Elder Scams}\label{sec:res-detect}

Our RQ2 analysis showed that older adults often reject relatives' warnings and continue to trust scammers. To examine relatives' roles in fraud detection and narration, we focused on who discovered the scam in our Reddit dataset and relatives' accounts in our Baidu Tieba dataset.

\textbf{Family detection on Reddit}: We analyzed 13,082 Reddit posts (multiple detectors allowed) and found that 74.4\% of posts identified family members as detectors. In contrast, 1,889 posts (14.4\%) identified older adults, 350 (2.7\%) friends or neighbors, and 300 (2.3\%) banks or institutions. In addition, 851 posts (6.5\%) did not specify who discovered the suspected scam. We observed that relatives uncovered these scams from incidental clues in daily life. For example, after calling an older adult's bank about a supposed fraud-department call, a user heard ``this is the first time I'm hearing this'' from staff. In another post about an older mother, one user learned of a scam when a sibling texted ``asking why she was trying to sell the car on Facebook'' about a listing unknown to the household. Prior studies of online posts also describe relatives and other contacts who recognized or suspected a scam that the targeted person had not: younger family members who identified fraud against seniors, such as unnoticed charges, in Chinese RedNote posts~\cite{CHI25-FamilySafeguard}; Reddit posters who recognized a scam that an older parent or grandparent had not~\cite{CCS25-Advice}; and, in Reddit posts about romance-scam targets of any age who denied the scam, posters who were typically family members, friends, neighbors, or other relations~\cite{CHI25-EmotionHelp}. Our findings further echo prior work that family members are the primary detectors in elder abuse scams. 

\textbf{Relatives' first-hand accounts on Baidu Tieba:} We examined kinship references in 2,262 first-hand posts coded as help seeking, verification seeking, complaints, or personal narratives, a subset of the 3,094 first-hand posts. Kinship terms alone do not establish narrator identity, so we describe relatives' accounts qualitatively, alongside prior research on Baidu Tieba~\cite{Lee21-TiebaFraud}. For instance, when a mother received health-product calls after a TV order, a user wrote: ``I asked for the company's address and name. After checking with the industry and commerce bureau, [I found that] both were false. But my mother had been brainwashed. No matter how I explained that they were scammers, she would not believe me.'' Similarly, a user visited their grandparents and discovered a diluted liquid. The user learned it was ``so-called cell regeneration water'' at 475 yuan per bottle. In addition, a user who bought basic phones for older adults found ``by chance while checking the bills'' that each phone had been charged 100 to 300 yuan.

\textbf{Summary \& Discussion of RQ3.} Our two datasets record different aspects of family involvement (Section~\ref{sec:corpus-annotation}), yet both show relatives spotting or narrating scams. In a different at-risk population, Zhang et al.~\cite{SP26-IntlStudents} found that international students sought scam advice from personal contacts while facing uncertainty and fear about formal reporting. U.S. consumer portals take third-party complaints, and securities rules permit trusted contacts~\cite{FTC-23,FINRA-SeniorRules}. In China, the police-run 96110 hotline and the National Anti-Fraud Center app take fraud reports and consultations from the public~\cite{WuhanPSB25-AntiFraudTools}. However, relatives without an account or contract authority still lack clear reporting channels across banks, phone carriers, and local regulators. Based on these findings, we provide the following recommendations: (1) \textbf{Banks, phone carriers, and regulators should allow relatives to report fraud}, even when older account holders refuse to complain. When family members report abnormal transfers, fraudulent sales, or unusual phone bills, staff should verify the report without granting relatives account control. For instance, banks can route family flags to internal fraud review or temporary holds~\cite{FINRA-SeniorRules} while keeping control with the older account holder.  (2) \textbf{Consumer agencies, carriers, and banks should provide public lookup tools} to allow family members to verify suspicious transaction claims before more funds are transferred out. Lookup pages for business registrations, caller numbers, and official bank numbers can help verify callers or sellers. They could also offer a live Q\&A session for relatives to ask questions and authorize transactions. 

\section{RQ4: Family Protective actions and Remedy Channels}\label{sec:res-family}

Families described protective actions in response to suspected scams (RQ3).
In this section, we examine how recorded older-adult responses relate to reported family security work (Section~\ref{sec:res-security-work}). We also evaluate remedy channels and recovery outcomes (Section~\ref{sec:res-remedy}).

\subsection{Family  protective actions and Older-Adult Stance}
\label{sec:res-security-work}

\textbf{Household security measures:} We analyzed household protective measures across the 13,082 posts in our Reddit dataset. Posts mentioned formal reports or legal action ($n= 2,006$), locking accounts or credentials ($n= 1,344$), monitoring phone or network communications ($n= 785$), physical intervention in transactions ($n= 442$), and taking over finances ($n= 436$). Users were also asking how to protect older adults who rejected warnings. For example, one user wrote that a relative ``just laughs it off and says it's fun to talk'' when warned, leaving the user asking how to take her phone away. Another user pursued power of attorney and monitored social media after a relative ignored warnings: ``She'll act like she believes us and then go right back to the scammer''.

\textbf{Stance and family security actions:}
We compared continued-belief posts ($n= 563$) in Tieba against the remaining 3,370 posts. Those posts include general warnings and retold accounts without necessarily recording warning acceptance. Warning others appeared in 21.5\% of continued-belief posts, formal reports or legal actions in 11.9\%, and recovery efforts in 4.6\%. In the remaining posts, these appeared in 37.8\%, 27.5\%, and 24.8\%, respectively. These rates reflect cross-sectional associations.
We also analyzed $ n=3,979$ Reddit posts with coded older-adult stances. These posts comprised resistance or continued involvement in the Persistence group ($ n=2,705$) and warning acceptance in the Acceptance group ($ n=1,274$). Family security action appeared in 39.6\% and 56.8\% of these groups, respectively (odds ratio 0.50). Institutional contact rates remained similar at 59.2\% and 63.0\%, respectively (odds ratio 0.85). When the older adult accepted that they had been scammed, the family locked down accounts (changed passwords, turned on two-factor authentication, froze accounts) in 33.3\% of posts, against 8.9\% when the older adult resisted.
Relatives monitored communications in 17.4\% of posts ($n= 470$) under resistance or continued involvement, versus 6.9\% ($n= 88$) under warning acceptance. Relatives also took over finances in 7.2\% of posts ($n= 196$) under resistance or continued involvement versus 4.0\% ($n= 51$) under warning acceptance. In one post, a relative wrote: ``I took control of my relatives accounts and the scammer began to walk them through how to wrestle control back from me.'' 

\subsection{Remedy Channels and Recovery Outcomes}\label{sec:res-remedy}

\begin{figure}[!htb]
  \centering
  \includegraphics[width=.83\columnwidth]{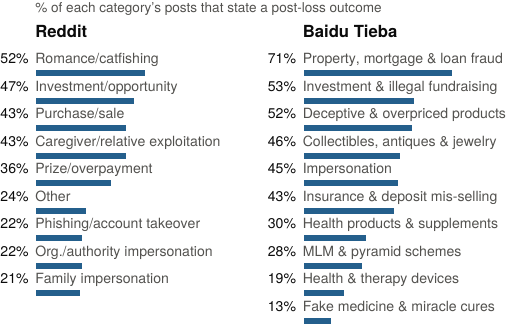}
  \caption{Outcome-stating posts by scam category.  Note: We showed ten largest categories for Tieba.}
  \label{fig:outcome-reporting}
\end{figure}

\textbf{Post-loss recovery rates:} Post-loss recovery remained rare in both datasets. Out of 13,082 Reddit posts, 4,254 (32.5\%) stated a post-loss recovery outcome (``recovered'', ``partially recovered'', or ``unrecovered''); of these, 233 (5.5\%) reported full ($n=83$) or partial ($n=150$) recovery. Out of 3,933 Tieba posts, 1,485 (37.8\%) stated a recovery outcome, and 140 (9.4\%) of those reported full or partial recovery. Figure~\ref{fig:outcome-reporting} breaks the 32.5\% and 37.8\% down by scam category. Among outcome-stating Reddit posts, 158 of 1,424 mentioning banks or payment institutions (11.1\%) reported recovery, versus 39 of 1,764 mentioning no institution (2.2\%). On Tieba, a reposted news report described a 78-year-old whose bank deposit had become an insurance policy; three trips to the bank failed, and the refund came only after a journalist appeared in the lobby.

A Tieba post illustrated the difficulty of documenting a suspected fraud after a purchase. An older relative had bought a phone advertised as new, but a family member suspected from its serial number that it was a replacement device. By then, the seven-day return window had passed, and the seller challenged the family to obtain an independent inspection. When the International Mobile Equipment Identity (IMEI) lookup services they contacted could not check the device, the poster turned to the forum to ask what they could do next. 
Together, such accounts illustrate why post-loss recovery is so rare. More specifically, by the time relatives detect a deceptive sale or transfer, dispute windows often have already been closed. This makes any pre-loss interception critical.

\begin{figure}[t]
  \centering
  \includegraphics[width=0.79\columnwidth]{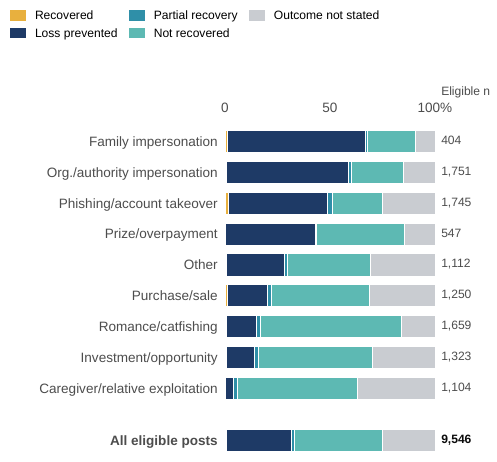}
  \caption{Reddit recovery outcomes ($n=9{,}546$ recovery-coded posts) in overlapping categories ($\geq40$ posts) by joint prevention or recovery percentage.}
  \label{fig:reddit-outcome}
\end{figure}

\textbf{Prevention and recovery by scam category:} We compared prevention and recovery across Reddit scam categories. Figure~\ref{fig:reddit-outcome} shows categories with at least 40 posts among the $n= 9,546$ recovery-coded Reddit posts. Across these posts, 30.6\% described pre-loss prevention, 42.1\% reported unrecovered losses, 24.9\% left outcomes unstated, and 2.4\% reported recovery. No scam category exceeded 4.0\% recovery. Pre-loss prevention was reported more often than post-loss recovery. Pre-loss prevention reached 65.6\% for family impersonation, 57.7\% for organization or authority impersonation, and 46.9\% for phishing scams. Pre-loss prevention dropped to 14.4\% in romance, 13.8\% in investment, and 3.8\% in caregiver or relative exploitation scams, where unrecovered losses reached 67.2\%, 54.4\%, and 57.2\%, respectively. User accounts illustrated early interception versus repeated extraction. In a prevented grandparent scam, one user wrote that an uncle ``kept the guy on the phone before telling him he'd never had children''. Another user noted a relative was ``currently trying to wire \$40,000 to his current scammer, and the banks are actually stopping him''; two banks said those policies had been in place for only a few months. In contrast, one user wrote that a relative spent seven years ``sending him money because she believed he was real'' and ``was kicked out of her home because she gave away all her money''.

\begin{figure}[!htb]\centering\includegraphics[width=.9\columnwidth]{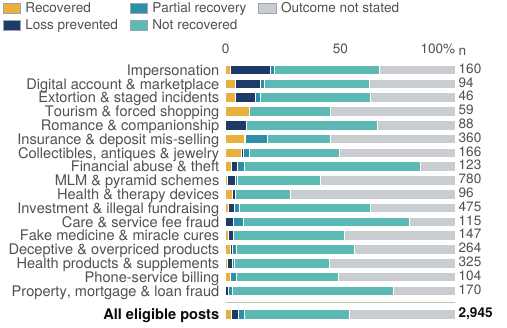}
\caption{17-category Tieba recovery outcomes ($\geq40$ recovery-coded posts, $n=2{,}945$, $n=988$ N/A excluded).}
\label{fig:tboutcome}\end{figure}

\textbf{Cross-category outcome distribution:} Figure~\ref{fig:tboutcome} shows outcomes for 17 scam categories with at least forty posts among all $n= 2,945$ recovery-coded Tieba posts (right-hand $n$ in the figure gives category denominators; unverified claims are coded as ``Outcome not stated''). Across these posts, 71 (2.4\%) reported full recovery, 69 (2.3\%) partial recovery, and 95 (3.2\%) pre-loss prevention. In 16 out of 17 categories, at least 80\% of posts either reported unrecovered losses or left outcomes unstated. Impersonation scams were the exception due to pre-loss prevention. In a news report reposted on Tieba about a 12,280 yuan water-treatment device advertised to treat diseases, a relative stated, ``They have already refunded me and taken the product back,'' after family negotiations. Family conflict persisted after the refund, as the older adult planned to buy another device.

\begin{figure*}[!t]\centering\includegraphics[width=.88\textwidth]{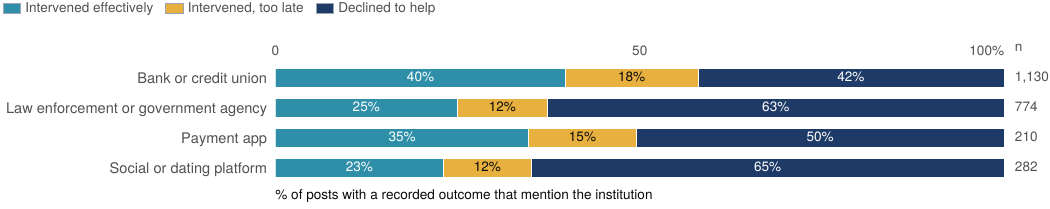}
\caption{Post-level institutional outcomes on Reddit across outcome-coded posts mentioning each institution (row $n$).}
\label{fig:instoutcomes}\end{figure*}

\textbf{Institutional contacts and outcomes:} Reddit posts most frequently mentioned banks and credit unions ($n= 2,413$), social media and dating platforms ($n= 2,024$), and law enforcement or government agencies ($n= 1,557$). Posts also mentioned payment applications (n=804), postal services ($n= 445$), mobile operators ($n= 429$), and gift card merchants ($n= 202$). Our institutional-response coding achieved 0.81 micro-$F_1$ on the development reference set (Table~\ref{tab:reddit-open-coding}). Institutional outcomes appeared in 50\% (774 out of 1,557) of law enforcement or government posts, 47\% (1,130 out of 2,413) of bank posts, and 14\% (282 out of 2,024) of social or dating platform posts. Figure~\ref{fig:instoutcomes} shows outcomes across these subsets. Labels reflect whole-post outcomes rather than per-institution ratings. Multi-institution posts therefore count in each row. Among outcome-coded posts, 63\% mentioning law enforcement or government agencies were coded as ``Declined to help'' and 25\% as ``Intervened effectively''; 65\% mentioning social or dating platforms as ``Declined to help''; and 42\% mentioning banks as ``Declined to help'', 18\% as ``Intervened, too late'', and 40\% as ``Intervened effectively''. Posts described gaps in institutional responses to victim-authorized transfers. After their mother in her 80s wired cash, an adult child wrote: ``The police said there was nothing they could do and no one stopped her or even questioned her.''  This pattern echoes prior work, in which banks and payment services offered little recourse once the victim had authorized the transfer~\cite{SOUPS25-Oak,CCS25-Advice} and police could do little for victims who reported~\cite{SOUPS25-Oak,SP24-PigButcher}.

\textbf{Recovery across institutional channels:} In our Tieba dataset, 1,485 posts stated a post-loss recovery outcome. Among them, 42.0\% (50 out of 119) of posts mentioning insurers and 36.4\% (20 out of 55) of posts mentioning banks reported full or partial recovery, versus 4.5\% (15 out of 330) of posts mentioning police. Regulated counterparties hold funds under identifiable contracts, such as insurance policies or bank deposits. Some posters reported that police classified voluntary transfers as civil economic disputes rather than criminal fraud. Users also described platform and telecom responses. One user reported an account to the platform after older relatives became obsessed with livestream hosts who ``call their fans Dad and Mom'', noting the platform only stopped recommending the account ``as if there were no penalty''. Another user wrote that a carrier claimed charges on a grandfather's feature phone came ``from his own use of game apps'' and then froze its billing function.

\textbf{Recovery by asset type:} Table~\ref{tab:asset-recovery} shows post-level recovery rates by asset type. ``Outcome stated'' counts recovered, partially recovered, or unrecovered posts, and ``Recovered'' gives recovery counts (\%). In outcome-stating Reddit posts, recovery reached 9.5\% for bank account access and 6.5\% for wire transfers. Recovery dropped to 4.3\% for cryptocurrency, 4.0\% for gift cards, and 3.5\% for cash. In outcome-stating Tieba posts, recovery reached 17.0\% for formal contracts or debt obligations versus 9.6\% for money transfers. Across datasets, recovery may reflect how reversible the payment channel or contract was.

\begin{table}[t]
\centering
\caption{Post-level full or partial recovery by overlapping asset types.}
\label{tab:asset-recovery}
\footnotesize
\setlength{\tabcolsep}{2pt}
\begin{tabular}{@{}llrrr@{}}\toprule
Platform & Asset & Posts & Outcome stated & Recovered \\ \midrule
Reddit & Bank account access & 1,292 & 746 & 71 (9.5\%) \\
Reddit & Wire transfer & 529 & 413 & 27 (6.5\%) \\
Reddit & Personal info only & 1,057 & 137 & 8 (5.8\%) \\
Reddit & Other & 2,267 & 1,332 & 68 (5.1\%) \\
Reddit & Cryptocurrency & 551 & 398 & 17 (4.3\%) \\
Reddit & Gift cards & 602 & 479 & 19 (4.0\%) \\
Reddit & Cash & 1,950 & 1,478 & 51 (3.5\%) \\
\midrule
Tieba & Data & 147 & 55 & 7 (12.7\%) \\
Tieba & Contract or Debt Obligation & 643 & 347 & 59 (17.0\%) \\
Tieba & Money & 3,229 & 1,407 & 135 (9.6\%) \\
Tieba & Labor & 327 & 103 & 4 (3.9\%) \\
Tieba & Health & 794 & 192 & 8 (4.2\%) \\
Tieba & Property & 107 & 82 & 4 (4.9\%) \\
\bottomrule
\end{tabular}
\end{table}

\textbf{Summary \& Discussion of RQ4.} We found family protective actions vary with older adults' response to warnings. Tieba posts recording continued belief rarely mention recovery efforts. On Reddit, pre-loss prevention is most frequent within sudden impersonation or phishing, and in outcome-stating posts, cash, gift-card, or crypto losses are rarely recovered. Based on these findings, we provide the following recommendations: (1) \textbf{Banks, payment apps, and stores should add checks before hard-to-reverse payments}. This applies especially to large withdrawals or gift-card and crypto purchases. Staff and apps can show warnings and hold transactions for review. For example, a sudden large transfer to a new account, which prior work asks banks to flag~\cite{SP24-PigButcher}, could wait for approval from a family member the customer chose in advance~\cite{CHI25-FamilySafeguard}. (2) \textbf{Banks, consumer agencies, and police should jointly review transfers approved by deceived customers}, instead of acting alone. Agarwal et al.~\cite{Agarwal25-HiMumDad} similarly recommend coordination among messaging services, mobile operators, and financial institutions in response to family-impersonation SMS scams. Shared complaint rules can include family reports~\cite{FTC-23,CHI25-FamilySafeguard} in fraud investigations rather than dismissing reported transfers as civil disputes. (3) \textbf{Online platforms should act promptly on reported scam accounts}, not just stop recommending them. Li et al.~\cite{NDSS24-CommentScams} found that YouTube scam replies could leave notifications after deletion and lead users to private messaging. Platforms should review reported accounts together with their linked contact details and act on confirmed scam activity.\\

\subsection{Importance of This Work}
Our findings reveal a critical gap in protecting older adults from continued financial exploitation. While family members often recognize suspected fraud (RQ3), older adults may continue trusting scammers despite warnings (RQ2), making it difficult for relatives to prevent repeated fund extraction and the resulting financial losses, debt, and family distress (RQ1). 
For example, families often struggle to lock down the victim’s accounts because the elder victim persistently believes they are not being scammed. 
At the same time, institutions may provide limited assistance when elder victims voluntarily authorize fraudulent transactions (RQ4).
Together, these findings highlight the gap between early fraud detection by family members and effective intervention to prevent further financial losses. Our work emphasizes the need for stronger fraud prevention and intervention mechanisms that address these challenges.

Existing safeguards often depend on account holders recognizing fraud and seeking assistance themselves, an assumption that may not hold when elder victims continue trusting scammers. 
Additionally, we found that law enforcement agencies declined to help 63\% of the time; therefore, we echo the need for law enforcement agencies to create specialized elder fraud units and partner with banks to freeze suspicious transactions before older adults lose their savings.
Furthermore, accessible third-party reporting channels, coordinated institutional responses, and independently reviewed protective measures could help prevent continued financial exploitation.
Therefore, this work offers insights for researchers and policymakers to investigate and develop frameworks, such as clinics, that could help at-risk elders navigate these difficult situations.

%% file: Limitations.tex
\section{Limitations \& Future Work} \label{Limit}

Our corpora differ in time spans, discussion formats, and codebooks, so measures are not directly comparable. Neither of our corpora tracks the same families over time. Restricting Tieba ($n= 3{,}933$) to May 2021--June 2026 ($n= 939$) shifts repeated extraction from 69.8\% ($2{,}619/3{,}752$) to 64.0\% ($581/908$), continued belief from 41.9\% ($443/1{,}057$) to 40.0\% ($100/250$), and recovery from 9.4\% ($140/1{,}485$) to 9.9\% ($44/446$). Our Reddit corpus covers only English-language subreddits; hence, we could miss the broader ecosystem of elder abuse scams in other languages.
Longitudinal interviews can follow belief changes, intervention agreement, and institutional responses over time. Future work can evaluate institutionally reviewed temporary holds prompted by third-party reports at banks and carriers.

%% file: Summary.tex
\section{Conclusion}
Our analysis of 13,082 Reddit and 3,933 Baidu Tieba scam discussions reveals that elder scams frequently involve repeated or ongoing victimization (Reddit) and extraction of funds (Tieba), resulting in financial losses, borrowing, and household strain (RQ1). 
Resistance to warnings or continued involvement is more frequently reported in Reddit romance and investment posts than in phishing or authority-impersonation posts, while continued belief on Tieba is concentrated in multi-level marketing, investment, and health-product sales (RQ2).
Although relatives are the primary recorded detectors of suspected fraud on Reddit and frequently narrate cases on Tieba (RQ3), recognizing a scam does not necessarily enable them to prevent further losses. As our opening example shows, older adults may keep trusting scammers despite warnings, making family intervention difficult.
Our findings further show that account locks or credential changes were less common under resistance or continued involvement, while monitoring and financial takeovers were more common in stance-coded Reddit posts (RQ4).
Together, these findings highlight a critical gap between fraud detection and effective intervention, underscoring the need for stronger third-party reporting and autonomy-preserving safeguards against continued financial exploitation in elder abuse scams.

%% file: Appendix.tex
\appendices

\section{Ethics Considerations}
\label{ethics}
We only collect publicly available posts, and we do not attempt to de-anonymize the poster. Although the data are publicly available, we treat all content as sensitive because of re-identification and retraumatization risks. We used large language models (LLMs) to filter and annotate posts based on the procedure outlined in the Methodology section~\ref{sec:corpus-annotation}. For screening, coding, reference labeling, and benchmarking, we sent post titles and bodies to hosted models; some steps also included post identifiers, forum names, or comment threads. Reddit screening and coding text went through OpenRouter’s routing service. Posts from both platforms also went to OpenAI and Anthropic models. We added custom masks to cover up names, phone numbers, and links in the text so no private data reached the providers.

\section{Codebooks}
\label{app:codebooks}

Figure~\ref{fig:codebook-overview} illustrates example sub-themes (not exact codebook values) across the lifecycle rows named in parentheses after each heading, which subdivide the broad lifecycle stages described in the main text.

\subsection{Reddit Codebook (16 fields)}
\label{code-reddit}
Options below are separated by semicolons. N/A marks absence; ``Other'' is never used to mark absence.

\noindent\textbf{1. Category}  (one or two). Romance scam/catfishing; Family/relative impersonation; Organization/authority impersonation; Investment/opportunity scam; Caregiver/relative financial exploitation; Phishing/account takeover/malware; Purchase/sale scam; Prize/overpayment scam; Other; N/A.

\noindent\textbf{2. Lure/Need Exploited} (multi-select). Companionship/loneliness; Fear of loss/consequence; Financial windfall; Helping family in crisis; Trust in public figure; Cognitive vulnerability; Other; N/A.

\noindent\textbf{3. Persuasion Tactic} : Persuasion \& Trust) (multi-select). Urgency; Authority impersonation; Fear/threat; Romantic flattery; Reciprocity; Isolation tactic; Persistent contact; Coached deception; Other; N/A.

\noindent\textbf{4. Trust Cue}  (multi-select). Institutional mimicry; Real or stolen photos/videos; Relationship before ask; Real personal/account information used; Other; N/A.

\noindent\textbf{5. Contact-to-Commitment}  (multi-select). Immediate payment; Escalating contact; Long-term relationship; In-person/video meeting; Repeat payments; N/A.

\noindent\textbf{6. Asset Extracted}  (multi-select). Cash; Gift cards; Wire transfer; Bank account access; Cryptocurrency; Personal info only; Attempt only, no loss; Other; N/A.

\noindent\textbf{7. A. Who noticed}  (multi-select). Family member; Bank/institution; Elder themselves; Friend/neighbor; Other; N/A.

\noindent\textbf{8. B. Detection cue/evidence}  (multi-select). Behavior change; Financial anomaly; Story inconsistency; Suspicious activity/communications noticed; Direct disclosure; External check; Failed promise; Other; Not stated.

\noindent\textbf{9. Older Adult's Response to Warning}  (single-select rule; multiple returned labels retained). Accepted correction; Denied/defended scammer; Minimized concern; Continued contact; Anger/conflict with family; N/A.

\noindent\textbf{10. Family Security Work} (multi-select). Account lockdown; Filed report/sought legal action; Took over finances; Monitored communications; Physically intervened; N/A; Other.

\noindent\textbf{11. Repeat vs. One-Off} (single-select). Single incident; Repeated, same scammer; Repeated, different scammers; Ongoing at posting; N/A.

\noindent\textbf{12. A. Institution present} (multi-select). Bank/credit union; Phone carrier; Social media/dating app; Gift card retailer; Law enforcement/government agency; Payment app/conduit; Postal service; N/A.

\noindent\textbf{13. B. What happened}  (single-select). Intervened effectively; Intervened, too late; Declined to help; N/A.

\noindent\textbf{14. Non-Financial Harms} (Figure~\ref{fig:codebook-overview}: Aftermath \& Recovery) (multi-select). Emotional distress/social isolation; Damaged family relationships; Loss of independence; Caregiver burnout; Physical health decline; N/A; Other.

\noindent\textbf{15. Emotion Tag}  (multi-select). Anger/frustration; Fear/anxiety; Grief/heartbreak; Guilt/self-blame; Shock/confusion; Suspicion; Sympathy/concern for elder; Autonomy tension; Help-seeking/problem-solving; Helplessness/exhaustion; Public-warning intent; Relief/gratitude; N/A; Other.

\noindent\textbf{16. Loss Recovery Status} (single-select). No (Not Recovered); Unknown (Outcome Unstated); Partial; Prevented; Yes (Recovered); N/A.

\subsection{Tieba Codebook (12 dimensions)}
\label{code-tieba}
Dimension 5 is recorded as two columns (contact channel; commitment). Multi-select limits are given per dimension.

\noindent\textbf{1. Scam Relevance}  (single-select). Yes (Scam-Relevant); Unknown (Undetermined); No (Not a Scam).

\noindent\textbf{2. Purpose of Post}  (single-select). Help \& Advice Seeking; Public Warning \& Education; Complaint \& Grievance; News \& Media Exposure; Verification \& Appraisal Seeking; Personal Narrative \& Testimony; Promotion \& Testimonial; Public Appeal \& Mobilization; Discussion \& Commentary.

\noindent\textbf{3. Category of Scam} (single or up to two). MLM \& Pyramid Schemes; Health Product \& Supplement Sales; Investment \& Illegal Fundraising; Deceptive \& Overpriced Product Sales; Insurance \& Deposit Mis-selling; Fake Medicine \& Miracle-Cure Treatment; Collectibles, Antiques \& Jewelry Scams; Health \& Therapy Device Scams; Telecom \& Phone-Service Billing Scams; Impersonation Scams; Property, Mortgage \& Loan Fraud; Tourism \& Forced-Shopping Scams; Romance \& Companionship Scams; Prize \& Lucky-Draw Scams; Elder Financial Abuse \& Theft; Extortion \& Staged-Incident Coercion; Fee-for-Service \& Care Fraud; Superstition \& Religious Scams; Digital Account, Data \& Marketplace Fraud; Other / Mixed Confidence Scams; N/A. Figure~\ref{fig:tbtrends} groups these categories into four clusters (with Other / Mixed Confidence Scams and N/A belonging to no cluster, and a post counting toward every cluster that contains one of its categories): Finance (Investment \& Illegal Fundraising; MLM \& Pyramid Schemes; Insurance \& Deposit Mis-selling; Property, Mortgage \& Loan Fraud; Elder Financial Abuse \& Theft), Consumer \& digital (Deceptive \& Overpriced Product Sales; Collectibles, Antiques \& Jewelry Scams; Telecom \& Phone-Service Billing Scams; Digital Account, Data \& Marketplace Fraud; Tourism \& Forced-Shopping Scams; Prize \& Lucky-Draw Scams), Health \& care (Fake Medicine \& Miracle-Cure Treatment; Health Product \& Supplement Sales; Fee-for-Service \& Care Fraud; Health \& Therapy Device Scams), and Social engineering \& coercion (Impersonation Scams; Romance \& Companionship Scams; Superstition \& Religious Scams; Extortion \& Staged-Incident Coercion).

\noindent\textbf{4. Promise / Need Exploited}  (single or multi-select (up to three)). Cure \& Disease Treatment; Health \& Wellness Enhancement; Investment Gain; Bargain Purchase \& Free Gifts; Income \& Business Opportunity; Collectible \& Valuables Appreciation; Safe Savings \& Principal Security; Device \& Telecom Deal; Pension \& Elder-Care Security; Prize, Subsidy \& Windfall; Companionship \& Romance; Insurance Protection \& Payout; Helping \& Family Emergency; Authority Compliance \& Threat Avoidance; Problem-Resolution \& Recovery Services; Spiritual \& Superstition; N/A.

\noindent\textbf{5. Contact-to-Commitment Route}  (two columns; A: one or two contact channels, B: one to three commitments; at most four combined). \emph{A. Contact channel:} Mass-Media Advertising \& Testimonial Content; Free-Gift, Prize \& Seminar Funnel; Family, Friend \& Acquaintance Channel; Provider, Store \& Existing-Account Channel; Bank \& Insurance Sales Channel; Street \& Door-to-Door Approach; Unspecified Sales Contact; Online Groups, Chat \& Platform Solicitation; Offline Recruitment \& Investment Solicitation; Telemarketing \& Phone Contact; Tour \& Travel-Group Funnel; Impersonation \& Pretext Contact; E-commerce \& Remote Purchase Platform; Livestream \& Short-Video Commerce. \emph{B. Commitment:} Purchase; Investment \& Deposit; Contract, Policy, Loan \& Property Signing; Direct Payment \& Transfer; Engagement, Attendance \& Channel Migration; Joining, Membership \& Recruitment; Coerced, Hidden \& Unauthorized Extraction; Treatment, Regimen \& Ritual Uptake; Relationship \& Grooming Commitment; Travel \& Inspection Trip; Access \& Disclosure; Contemplated / No Commitment; N/A.

\noindent\textbf{6. Asset Extracted}  (single or multi-select (up to two)). Money; Health; Property; Contract or Debt Obligation; Labor; Data; N/A.

\noindent\textbf{7. Red Flags} (multi-select; max five). False Health Claims \& Endangerment; Unrealistic Promises \& Scheme Mechanics; Faked Identity, Authority \& Proof; Product \& Legitimacy Deception; Pressure \& Escalating Demands; Covert Terms \& Charges; Non-Delivery, Withdrawal Failure \& Vanishing; Grooming, Isolation \& Bait; Targeting \& Access Exploitation; N/A.

\noindent\textbf{8. Response \& Protective Action}  (multi-select (up to four)). Seek Help \& Advice; Warn \& Educate Others; Verify \& Investigate; Report \& Legal Action; Seek Remedy \& Recover; Dissuade \& Confront Elder; Take Control \& Secure; Public Appeal \& Collective Action; Accept Comply; Continue Persist In Belief; Promote Advocate Scheme; Reject Refuse; Stop Disengage; Distress \& Regret; Other Respond; N/A.

\noindent\textbf{9. One-Off vs. Repeated} (single-select). One-Off; Repeated; N/A.

\noindent\textbf{10. Party Intervened} (single or up to two; N/A when no third party appears). Police Response; Government, Judicial, Regulatory \& Legal Aid; Bank Intervention; Insurer Response; Seller / Company Response; Telecom Carrier Response; Platform Moderation; Media Exposure \& Investigation; Medical \& Care Institutions; N/A.

\noindent\textbf{11. Emotional Frame}  (one or two dominant frames; never more than two). Anger, Outrage \& Injustice; Protective Concern \& Sympathy; Warning \& Caution; Frustration \& Exasperation; Helplessness \& Desperation; Suspicion \& Distrust; Fear, Alarm \& Urgency; Promotional Appeal; Grief, Hurt \& Regret; Caregiver Strain \& Distress; Ridicule \& Sarcasm; Relief \& Vindication; Help-Seeking.

\noindent\textbf{12. Loss Recovery Status} (single-select). No (Not Recovered); Unknown (Outcome Unstated); Partial; Prevented; Yes (Recovered); N/A.

\noindent\emph{Auxiliary narration field (retained from the initial codebook for corpus stratification; not scored in the coding benchmark):} Poster Is First-Hand Participant (binary single-select). Yes (First-Hand Participant); No (Third-Party Retelling).

\section{Coding and Screening Benchmarks}
\label{app:benchmarks}
\begin{table}[!h]
\centering
\caption{Baidu Tieba scam-relevance screening benchmark (\texttt{-avg}: mean across all reasoning tiers).}
\label{tab:bench-tieba-tp}
\footnotesize
\begin{tabular}{@{}lrrrr@{}}\toprule
Configuration & Acc & P & R & $F_1$ \\\midrule
gpt-5.5-xhigh & 0.9250 & 0.5524 & \textbf{0.9364} & 0.6949 \\
kimik25-medium & 0.9238 & 0.5483 & \textbf{0.9342} & 0.6910 \\
bench-sonnet-max & 0.9254 & 0.5541 & \textbf{0.9320} & 0.6950 \\
dsv32-high & 0.8996 & 0.4743 & \textbf{0.9320} & 0.6287 \\
qwen37flash-medium & 0.9228 & 0.5450 & \textbf{0.9298} & 0.6872 \\
kimik25-high & 0.9254 & 0.5544 & \textbf{0.9276} & 0.6940 \\
dsv32-medium & 0.9016 & 0.4795 & \textbf{0.9254} & 0.6317 \\
grok45-high & 0.9242 & 0.5502 & \textbf{0.9254} & 0.6901 \\
bench-sonnet-avg & 0.9309 & 0.5774 & \textbf{0.9079} & 0.7056 \\
gpt-5.6-sol-avg & 0.9280 & 0.5663 & \textbf{0.9041} & 0.6963 \\
dsv4flash-avg & 0.9280 & 0.5663 & \textbf{0.8992} & 0.6949 \\
qwen37max-avg & 0.9319 & 0.5824 & \textbf{0.8958} & 0.7059 \\
gemini31pro-high & 0.9312 & 0.5800 & \textbf{0.8904} & 0.7024 \\
qwen37plus-avg & 0.9329 & 0.5878 & \textbf{0.8849} & 0.7064 \\
dsv4pro-avg & 0.9338 & 0.5926 & \textbf{0.8772} & 0.7074 \\
glm52-avg & 0.9347 & 0.5973 & \textbf{0.8728} & 0.7092 \\
roberta-wwm-ext-ft & 0.9548 & 0.7106 & \textbf{0.8509} & 0.7745 \\
macbert-base-ft & 0.9558 & 0.7238 & \textbf{0.8333} & 0.7747 \\
bert-base-chinese-ft & 0.9596 & 0.7900 & \textbf{0.7588} & 0.7740 \\
tfidf-logreg & 0.9308 & 0.5968 & \textbf{0.7434} & 0.6621 \\
bow-logreg & 0.9252 & 0.5737 & \textbf{0.6996} & 0.6304 \\
\bottomrule\end{tabular}
\end{table}
\begin{table}[!t]
\centering
\caption{Reddit scam-relevance screening benchmark (\texttt{-avg}: mean across all reasoning tiers).}
\label{tab:bench-reddit-tp}
\footnotesize
\begin{tabular}{@{}lrrrr@{}}\toprule
Configuration & Acc & P & R & $F_1$ \\\midrule
qwen37flash-medium & 0.9024 & 0.5914 & \textbf{0.9374} & 0.7252 \\
qwen37flash-high & 0.9020 & 0.5903 & \textbf{0.9374} & 0.7244 \\
qwen38-27b-medium & 0.9040 & 0.5965 & \textbf{0.9316} & 0.7273 \\
gpt-5.5-low & 0.9030 & 0.5940 & \textbf{0.9287} & 0.7246 \\
qwen38max-medium & 0.9106 & 0.6172 & \textbf{0.9199} & 0.7387 \\
gpt-5.5-avg & 0.9034 & 0.5965 & \textbf{0.9189} & 0.7233 \\
bench-sonnet-avg & 0.9077 & 0.6090 & \textbf{0.9170} & 0.7319 \\
gpt-5.6-sol-avg & 0.9042 & 0.5991 & \textbf{0.9164} & 0.7245 \\
dsv4pro-medium & 0.9054 & 0.6025 & \textbf{0.9156} & 0.7267 \\
dsv4flash-avg & 0.9109 & 0.6190 & \textbf{0.9149} & 0.7383 \\
gpt-5.6-luna-avg & 0.9056 & 0.6032 & \textbf{0.9147} & 0.7270 \\
deberta-v3-base-ft & 0.9348 & 0.7030 & \textbf{0.9098} & 0.7931 \\
gpt-5.4-mini-avg & 0.9014 & 0.5920 & \textbf{0.9094} & 0.7171 \\
gpt-5.6-terra-avg & 0.9053 & 0.6040 & \textbf{0.9025} & 0.7237 \\
bench-haiku-avg & 0.9126 & 0.6266 & \textbf{0.9010} & 0.7391 \\
bert-base-uncased-ft & 0.9394 & 0.7388 & \textbf{0.8646} & 0.7968 \\
roberta-base-ft & 0.9414 & 0.7613 & \textbf{0.8355} & 0.7967 \\
tfidf-wordchar-logreg & 0.8858 & 0.5562 & \textbf{0.8355} & 0.6678 \\
bow-logreg & 0.8798 & 0.5530 & \textbf{0.6536} & 0.5991 \\
\bottomrule\end{tabular}
\end{table}

\begin{table}[!h]
{\centering
\caption{Reddit coding benchmark (micro-$F_1$ by codebook variable; \texttt{-medium}: medium across all reasoning tiers). Columns map to Appendix~B.1 fields: \emph{Cat}: Category; \emph{Lure}: Lure/Need Exploited; \emph{Tact}: Persuasion Tactic; \emph{Trust}: Trust Cue; \emph{Cont}: Contact-to-Commitment; \emph{Asset}: Asset Extracted; \emph{D: who}: Who Noticed; \emph{Resp}: Older Adult's Response to Warning; \emph{Sec}: Family Security Work;  }
\label{tab:reddit-coding-full}\label{tab:reddit-open-coding}
\fontsize{6.4bp}{7.2bp}\selectfont
\setlength{\tabcolsep}{0.45bp}
\newcommand{\redditLabel}[1]{{%
  \fontsize{5.75bp}{6.2bp}\selectfont\shortstack[c]{#1}}}
\newcommand{\redditHead}[2][0bp]{%
  \makebox[1.75em][c]{\kern#1\relax\redditLabel{#2}\kern\dimexpr-#1\relax}}
\begin{tabular*}{\columnwidth}{@{\extracolsep{\fill}}%
  >{\raggedright\arraybackslash}p{39bp}*{16}{r}@{}}
\toprule
\redditLabel{Configuration}
 & \redditHead[-1.4bp]{Cat.} & \redditHead[-1.4bp]{Lure}
 & \redditHead[-1.4bp]{Tact.} & \redditHead[-1.4bp]{Trust}
 & \redditHead[-0.6bp]{Cont.} & \redditHead[0.6bp]{Asset}
 & \redditHead[0.6bp]{D:\\who} & \redditHead[0.6bp]{Resp.}
 & \redditHead[0.5bp]{Sec.} & \redditHead{Rep.}
 & \redditHead[-0.6bp]{Inst.\\p} & \redditHead[-1.9bp]{Inst.\\o}
 & \redditHead[-4.25bp]{Harms} & \redditHead[-1.05bp]{Emot.}
 & \redditLabel{Recov.} & \redditLabel{Pooled} \\
\midrule
glm-5.3-flash & \textbf{0.81} & 0.72 & 0.62 & 0.59 & 0.63 & 0.59 & 0.78 & \textbf{0.72} & 0.54 & 0.68 & 0.70 & 0.81 & 0.70 & \textbf{0.70} & 0.61 & \textbf{0.68} \\
gpt-5.5-high & 0.78 & \textbf{0.75} & 0.66 & 0.63 & 0.64 & 0.64 & 0.79 & 0.68 & 0.52 & 0.68 & 0.62 & 0.83 & 0.56 & 0.66 & \textbf{0.71} & 0.67 \\
gpt-5.5-xhigh & 0.77 & 0.72 & \textbf{0.66} & 0.64 & 0.67 & 0.61 & 0.78 & 0.65 & 0.54 & 0.68 & 0.62 & 0.82 & 0.59 & 0.65 & 0.71 & 0.67 \\
gpt-5.5-low & 0.77 & 0.75 & 0.64 & \textbf{0.66} & 0.65 & 0.63 & 0.80 & 0.66 & 0.52 & 0.64 & 0.62 & 0.79 & 0.59 & 0.66 & 0.69 & 0.67 \\
gpt-5.5-medium & 0.77 & 0.74 & 0.64 & 0.62 & \textbf{0.68} & 0.60 & 0.81 & 0.65 & 0.53 & 0.67 & 0.64 & 0.80 & 0.59 & 0.64 & 0.68 & 0.66 \\
gpt-5.6-sol-medium & 0.76 & 0.72 & 0.65 & 0.53 & 0.62 & 0.52 & 0.77 & 0.66 & \textbf{0.55} & 0.68 & 0.70 & 0.82 & 0.64 & 0.63 & 0.70 & 0.66 \\
qwen3.8-27b & 0.78 & 0.70 & 0.59 & 0.58 & 0.63 & \textbf{0.66} & 0.75 & 0.70 & 0.51 & \textbf{0.69} & \textbf{0.71} & \textbf{0.86} & 0.65 & 0.63 & 0.52 & 0.66 \\
qwen3.8-flash & 0.79 & 0.71 & 0.61 & 0.59 & 0.64 & 0.63 & 0.78 & 0.59 & 0.46 & 0.69 & 0.63 & 0.83 & \textbf{0.70} & 0.67 & 0.59 & 0.66 \\
deepseek-v4-flash & 0.75 & 0.74 & 0.62 & 0.61 & 0.65 & 0.54 & 0.80 & 0.64 & 0.53 & 0.69 & 0.63 & 0.83 & 0.67 & 0.66 & 0.53 & 0.65 \\
qwen3.7-flash & 0.80 & 0.73 & 0.62 & 0.59 & 0.63 & 0.56 & \textbf{0.82} & 0.65 & 0.50 & 0.66 & 0.66 & 0.74 & 0.65 & 0.68 & 0.51 & 0.65 \\
\textit{GPT median} & 0.77 & 0.72 & 0.62 & 0.55 & 0.62 & 0.53 & 0.78 & 0.64 & 0.54 & 0.65 & 0.68 & 0.82 & 0.65 & 0.63 & 0.68 & 0.65 \\
kimi-k2.6 & 0.73 & 0.70 & 0.57 & 0.59 & 0.66 & 0.50 & 0.79 & 0.65 & 0.50 & 0.69 & 0.67 & 0.82 & 0.68 & 0.64 & 0.63 & 0.65 \\
deepseek-v4-pro & 0.79 & 0.69 & 0.58 & 0.57 & \textbf{0.68} & 0.54 & 0.81 & 0.66 & 0.49 & 0.62 & 0.67 & 0.81 & 0.68 & 0.63 & 0.55 & 0.65 \\
\addlinespace[2pt]\cmidrule(l{2pt}r{2pt}){1-17}\addlinespace[1pt]
bow-logreg & 0.61 & 0.45 & 0.33 & 0.28 & 0.37 & 0.22 & 0.69 & 0.48 & 0.41 & 0.41 & 0.49 & 0.78 & 0.52 & 0.51 & 0.31 & 0.46 \\
svd-mlp & 0.57 & 0.36 & 0.26 & 0.23 & 0.29 & 0.17 & 0.65 & 0.48 & 0.41 & 0.45 & 0.50 & 0.76 & 0.55 & 0.51 & 0.30 & 0.43 \\
tfidf-logreg & 0.59 & 0.25 & 0.18 & 0.13 & 0.23 & 0.05 & 0.73 & 0.42 & 0.35 & 0.45 & 0.52 & 0.76 & 0.51 & 0.49 & 0.30 & 0.40 \\
roberta-base-ft & 0.47 & 0.22 & 0.16 & 0.08 & 0.30 & 0.05 & 0.68 & 0.50 & 0.34 & 0.48 & 0.53 & 0.76 & 0.43 & 0.46 & 0.30 & 0.38 \\
bert-base-uncased-ft & 0.47 & 0.09 & 0.14 & 0.04 & 0.24 & 0.01 & 0.69 & 0.50 & 0.34 & 0.52 & 0.46 & 0.76 & 0.42 & 0.43 & 0.31 & 0.36 \\
distilbert-base-uncased-ft & 0.44 & 0.03 & 0.12 & 0.03 & 0.22 & 0.01 & 0.68 & 0.49 & 0.34 & 0.44 & 0.48 & 0.76 & 0.42 & 0.45 & 0.30 & 0.35 \\
\bottomrule
\end{tabular*}
}
\end{table}

\begin{table}[!h]
{\centering
\caption{Baidu Tieba coding benchmark (micro-$F_1$ by codebook variable).}
\label{tab:tieba-coding-full}\label{tab:tieba-coding-merged-deck}
\fontsize{7.0bp}{8.0bp}\selectfont\setlength{\tabcolsep}{0.4bp}
\begin{tabular*}{\columnwidth}{@{\extracolsep{\fill}}>{\raggedright\arraybackslash}p{49bp}rrrrrrrrrrrrr@{}}
\toprule
{\fontsize{6.0bp}{6.6bp}\selectfont\shortstack[r]{Configuration}} & {\fontsize{6.0bp}{6.6bp}\selectfont\shortstack[r]{Rele-\\vance}} & {\fontsize{6.0bp}{6.6bp}\selectfont\shortstack[r]{Pur-\\pose}} & {\fontsize{6.0bp}{6.6bp}\selectfont\shortstack[r]{Cat-\\egory}} & {\fontsize{6.0bp}{6.6bp}\selectfont\shortstack[r]{Prom-\\ise}} & {\fontsize{6.0bp}{6.6bp}\selectfont\shortstack[r]{Con-\\tact}} & {\fontsize{6.0bp}{6.6bp}\selectfont\shortstack[r]{Asset}} & {\fontsize{6.0bp}{6.6bp}\selectfont\shortstack[r]{Red\\flags}} & {\fontsize{6.0bp}{6.6bp}\selectfont\shortstack[r]{Re-\\sponse}} & {\fontsize{6.0bp}{6.6bp}\selectfont\shortstack[r]{Repeat}} & {\fontsize{6.0bp}{6.6bp}\selectfont\shortstack[r]{Party}} & {\fontsize{6.0bp}{6.6bp}\selectfont\shortstack[r]{Emo-\\tion}} & {\fontsize{6.0bp}{6.6bp}\selectfont\shortstack[r]{Re-\\covery}} & {\fontsize{6.0bp}{6.6bp}\selectfont\shortstack[r]{Pooled}} \\
\midrule
qwen3.8-flash & 0.87 & 0.71 & 0.82 & \textbf{0.79} & 0.71 & 0.82 & 0.62 & 0.60 & 0.78 & 0.80 & 0.61 & 0.64 & \textbf{0.71} \\
qwen3.7-flash & \textbf{0.89} & 0.77 & 0.83 & 0.75 & 0.70 & 0.80 & 0.62 & 0.59 & 0.69 & 0.79 & 0.60 & 0.65 & 0.70 \\
glm-5.3-flash & 0.88 & 0.73 & 0.87 & 0.75 & \textbf{0.72} & 0.81 & \textbf{0.67} & 0.53 & 0.77 & 0.80 & 0.58 & 0.61 & 0.70 \\
kimi-k2.6 & 0.87 & 0.75 & \textbf{0.88} & 0.77 & 0.70 & 0.79 & 0.63 & 0.60 & 0.62 & 0.77 & 0.58 & \textbf{0.65} & 0.70 \\
qwen3.8-27b & 0.88 & 0.74 & 0.80 & 0.72 & 0.69 & 0.79 & 0.61 & 0.58 & 0.74 & \textbf{0.81} & 0.57 & 0.54 & 0.69 \\
deepseek-v4-pro & 0.86 & 0.74 & 0.77 & 0.75 & 0.69 & 0.79 & 0.59 & 0.57 & 0.61 & 0.77 & 0.52 & 0.61 & 0.67 \\
tfidf-logreg & 0.86 & 0.67 & 0.76 & 0.70 & 0.61 & 0.78 & 0.61 & \textbf{0.63} & 0.79 & 0.67 & \textbf{0.61} & 0.54 & 0.67 \\
deepseek-v4-flash & 0.72 & \textbf{0.78} & 0.80 & 0.74 & 0.66 & 0.76 & 0.61 & 0.54 & 0.64 & 0.77 & 0.56 & 0.59 & 0.66 \\
svd-mlp & 0.85 & 0.68 & 0.76 & 0.70 & 0.64 & 0.79 & 0.55 & 0.62 & 0.80 & 0.53 & 0.58 & 0.58 & 0.66 \\
bow-logreg & 0.86 & 0.63 & 0.75 & 0.67 & 0.59 & 0.79 & 0.60 & 0.63 & \textbf{0.82} & 0.69 & 0.54 & 0.51 & 0.66 \\
roberta-wwm-ext-ft & 0.87 & 0.70 & 0.49 & 0.44 & 0.45 & 0.80 & 0.52 & 0.61 & 0.78 & 0.62 & 0.55 & 0.56 & 0.60 \\
bert-base-chinese-ft & 0.86 & 0.71 & 0.52 & 0.48 & 0.47 & 0.81 & 0.51 & 0.59 & 0.77 & 0.59 & 0.53 & 0.51 & 0.59 \\
macbert-base-ft & 0.86 & 0.71 & 0.50 & 0.45 & 0.44 & 0.78 & 0.52 & 0.60 & 0.78 & 0.60 & 0.54 & 0.56 & 0.59 \\
rbt3-ft & 0.88 & 0.65 & 0.43 & 0.39 & 0.37 & \textbf{0.82} & 0.42 & 0.58 & 0.74 & 0.57 & 0.51 & 0.55 & 0.55 \\
\bottomrule
\end{tabular*}
}
\end{table}

Tables~\ref{tab:bench-tieba-tp} and~\ref{tab:bench-reddit-tp} report the scam-relevance screening benchmarks. Each row is one configuration, and columns report accuracy, precision, recall (in bold), and $F_1$. Configurations are ranked by recall. In Table~\ref{tab:bench-reddit-tp}, ties are broken by precision. Table~\ref{tab:bench-tieba-tp} shows 46 configurations on the frozen Baidu Tieba set of $n=5{,}000$ posts ($n=456$ positive and $n=4{,}544$ negative). Table~\ref{tab:bench-reddit-tp} shows 50 configurations on the frozen Reddit set of $n=5{,}000$ posts ($n=687$ positive and $n=4{,}313$ negative). We used GPT-5.5 with xhigh reasoning for Baidu Tieba screening (0.936 recall, 0.552 precision, 0.695 $F_1$ on the reference set) and \texttt{qwen3.7-flash} with medium reasoning for Reddit (0.937 recall, 0.591 precision, 0.725 $F_1$).

Table~\ref{tab:tieba-coding-full} shows the Baidu Tieba coding benchmark across $n=201$ human reference posts ($n=227$ hand-coded minus $n=26$ worked examples) on 12 variables. Table~\ref{tab:reddit-coding-full} shows the retrospective Reddit coding benchmark across $n=161$ human reference posts ($n=200$ hand-coded minus $n=39$ worked examples) on 15 fields. Each cell reports development-set micro-$F_1$.
Local coding models were trained on 2,601 Reddit and 7,233 Baidu Tieba gold-labeled posts.
Bold values mark column maxima (rounded ties differ in unrounded values), and the GPT median row highlighted in italics in Table~\ref{tab:reddit-coding-full} summarizes 17 configurations, including those displayed.
We used \texttt{glm-5.3-flash} for the 13,082-post Reddit corpus and \texttt{qwen3.8-flash} for the Tieba labels before the manual label review of the final 3,933 posts described in Section~\ref{sec:corpus-annotation}.